\documentclass[twocolumn]{aastex631}

\shorttitle{SN\,1987A}
\shortauthors{Rosu et al.}
\graphicspath{{./}{figures/}}
\usepackage[referable]{threeparttablex}
  
\def\fds{\hbox{$.\!\!{''}$}}  
\def\fdsec{\hbox{$.\!\!^{\text{s}}$}}  
\usepackage{amsmath}
\DeclareMathOperator{\asinh}{asinh}
\usepackage[version=4]{mhchem}
\usepackage{bold-extra}

\begin{document}

\title{Hubble Space Telescope images of SN\,1987A:\\Evolution of the ejecta and the equatorial ring from 2009 to 2022}

\author[0000-0002-2461-6913]{Sophie Rosu}
\affiliation{Department of Physics, KTH Royal Institute of Technology, The Oskar Klein Centre, AlbaNova, SE-106 91 Stockholm, Sweden}
\author[0000-0003-0065-2933]{Josefin Larsson}
\affiliation{Department of Physics, KTH Royal Institute of Technology, The Oskar Klein Centre, AlbaNova, SE-106 91 Stockholm, Sweden}
\author[0000-0001-8532-3594]{Claes Fransson}
\affiliation{Department of Astronomy, Stockholm University, The Oskar Klein Centre, AlbaNova, SE-106 91 Stockholm, Sweden}
\author{Peter Challis}
\affiliation{Harvard-Smithsonian Center for Astrophysics, 60 Garden Street, MS-19, Cambridge, MA 02138, USA}
\author[0000-0002-5477-0217]{Tuomas Kangas}
\affiliation{Finnish Centre for Astronomy with ESO (FINCA), FI-20014 University of Turku, Finland}
\affiliation{Tuorla Observatory, Department of Physics and Astronomy, FI-20014 University of Turku, Finland}
\author[0000-0002-1966-3942]{Robert P. Kirshner}
\affiliation{Thirty Meter Telescope International Observatory, 100 West Walnut Street, Pasadena, California, USA 91124}
\author[0000-0002-7491-7052]{Stephen S. Lawrence}
\affiliation{Department of Physics and Astronomy, Hofstra University, Hempstead, NY 11549, USA}
\author[0000-0002-3664-8082]{Peter Lundqvist}
\affiliation{Department of Astronomy, Stockholm University, The Oskar Klein Centre, AlbaNova, SE-106 91 Stockholm, Sweden}
\author[0000-0002-5529-5593]{Mikako Matsuura}
\affiliation{Cardiff Hub for Astrophysics Research and Technology (CHART), School of Physics and Astronomy, Cardiff University, Queen's Buildings, The Parade, Cardiff CF24 3AA, UK}
\author[0000-0003-1546-6615]{Jesper Sollerman}
\affiliation{Department of Astronomy, Stockholm University, The Oskar Klein Centre, AlbaNova, SE-106 91 Stockholm, Sweden}
\author[0000-0003-1440-9897]{George Sonneborn}
\affiliation{Observational Cosmology Laboratory, Code 665, NASA Goddard Space Flight Center, Greenbelt, MD 20771, USA}
\author[0000-0002-7746-8512]{Linda Tenhu}
\affiliation{Department of Physics, KTH Royal Institute of Technology, The Oskar Klein Centre, AlbaNova, SE-106 91 Stockholm, Sweden}

\begin{abstract}
Supernova (SN) 1987A offers a unique opportunity to study how a spatially resolved SN evolves into a young supernova remnant (SNR). We present and analyze Hubble Space Telescope (HST) imaging observations of SN\,1987A obtained in 2022 and compare them with HST observations from 2009 to 2021. These observations allow us to follow the evolution of the equatorial ring (ER), the rapidly expanding ejecta, and emission from the center over a wide range in wavelength from 2000 to 11\,000\,\AA. The ER has continued to fade since it reached its maximum $\sim8200$ days after the explosion. In contrast, the ejecta brightened until day $\sim11\,000$ before their emission levelled off; the west side brightened more than the east side, which we attribute to the stronger X-ray emission by the ER on that side. 
The asymmetric ejecta expand homologously in all filters, which are dominated by various emission lines from hydrogen, calcium, and iron. From this overall similarity, we infer the ejecta are chemically well-mixed on large scales. The exception is the diffuse morphology observed in the UV filters dominated by emission from the Mg\,\textsc{ii} resonance lines that get scattered before escaping. The 2022 observations do not show any sign of the compact object that was inferred from highly-ionized emission near the remnant's center observed with JWST. 
We determine an upper limit on the flux from a compact central source in the [O\,\textsc{iii}] HST image. The non-detection of this line indicates that the S and Ar lines observed with JWST originate from the O free inner Si -- S -- Ar rich zone and/or that the observed [O\,\textsc{iii}] flux is strongly affected by dust scattering.
\end{abstract}


\section{Introduction\label{sect:introduction}}
Supernova (SN) 1987A in the Large Magellanic Cloud (LMC), discovered on 1987 February 23, is the closest naked-eye SN since Kepler's SN in 1604 \citep[see][for reviews]{arnett89, mccray93, mccray16}. This unique event became one of the most deeply studied astronomical objects in the Universe outside of our solar system, from ground to space, at all wavelengths. Its evolution from supernova to supernova remnant has notably been followed by the Hubble Space Telescope (HST) since its launch in 1990 with an excellent spatial resolution \citep[see e.g.,][and references therein]{larsson19a}. 

\begin{figure*}
\centering
\includegraphics[clip=true, trim=70 60 55 65,width=0.75\linewidth]{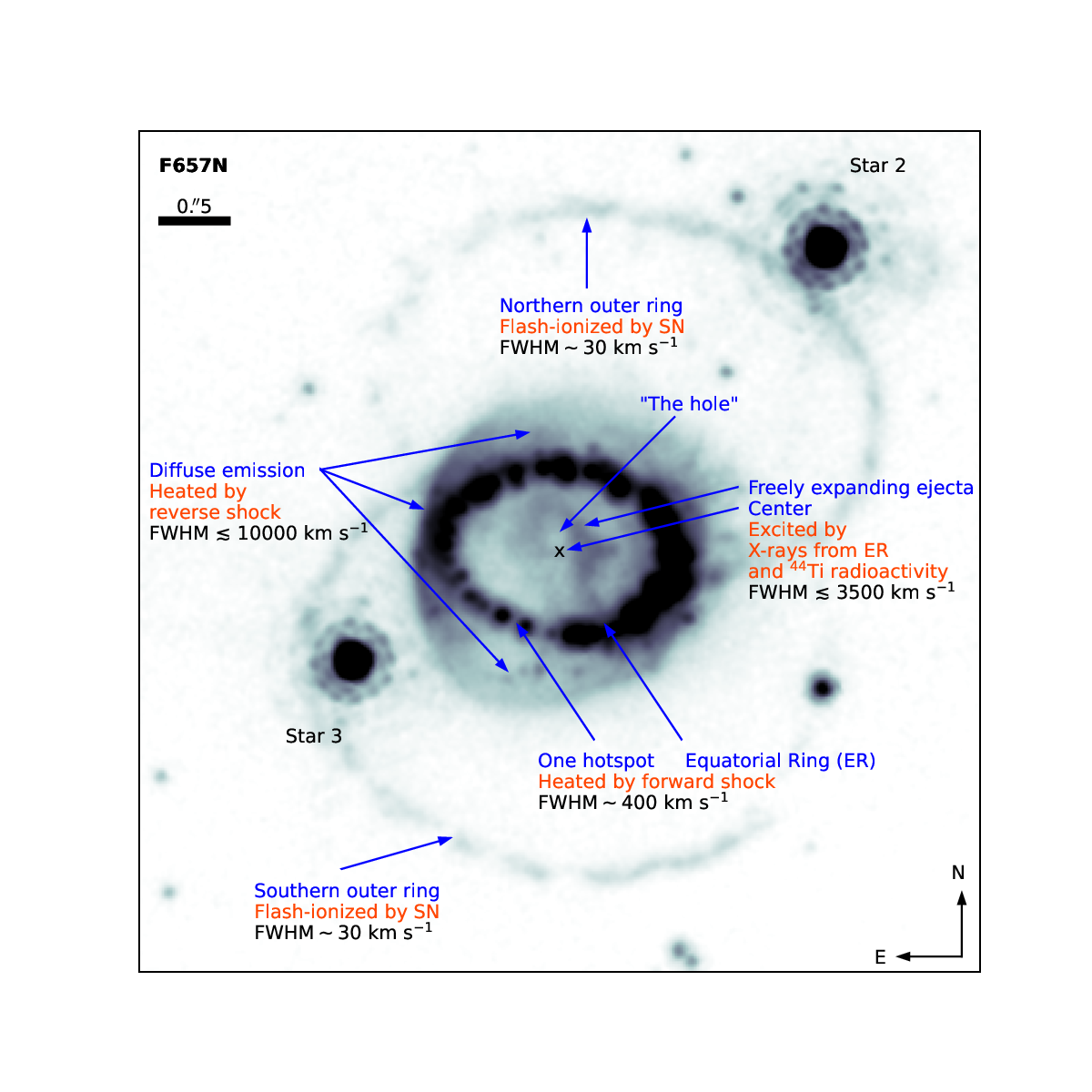}
\caption{HST/WFC3 2022 image of SN\,1987A in the F657N filter with labels showing the main emission components (dark blue) and corresponding processes responsible for these emissions (orange). The FWHM of the optical emission lines are indicated in black (from \citealt{groningsson08,tziamtzis11,fransson13,larsson19a}). 
The image was scaled by an $\asinh$ function to highlight the weak emission outside the ER. The field of view is $6\fds0 \times 6\fds0$. The cross represents the geometric center of the ER \citep{alp18}.}
\label{fig:general}
\end{figure*}

The resolved view offered by the HST images taken with the Wide Field Camera 3 (WFC3) shows the ejecta expanding into the circumstellar triple-ring nebula (see Fig.\,\ref{fig:general}). The inner equatorial ring (ER) has a radius of $\sim 0\fds8$, while the fainter outer rings have radii almost three times larger and are offset from the equatorial plane \citep[][their figure~5]{tziamtzis11}. The three rings are inclined by an angle ranging between $38^\circ$ and $45^\circ$ \citep{tziamtzis11}, $43^\circ$ being the most probable value for the ER inclination, with the northern part being the near side \citep{panagia91, plait95, sugerman05}. The pre-shock expansion velocities of the rings imply they were ejected $\sim20\,000$ years before the explosion \citep{crotts00}. The formation of the triple ring was likely connected to a binary merger \citep{morris07,morris09}. 

Figure~\ref{fig:general} illustrates the asymmetric, dense inner ejecta which are expanding freely inside the ER, as well as the shock interaction between the fast outer ejecta and the circumstellar medium (CSM). 
The key properties of the different emission components are summarized in Fig.\,\ref{fig:general}. The velocity of the freely expanding ejecta ($v_{\rm ej}$) is given by 
\begin{equation}
\label{eq:ejectavel}
v_{\rm ej} = 858 \left( \frac{t}{10\,000\ {\rm days}}\right)^{-1}   \left( \frac{d}{0\farcs{1}} \right) \rm{km\,s^{-1}}, 
\end{equation}
where $t$ is the time since explosion and $d$ is the angular distance from the geometric center (marked with a cross in Fig.\,\ref{fig:general}), assuming a distance to the LMC of 49.59\,kpc \citep{pietrzynski19}. The most recent epoch considered in this paper is $t \sim 13\,000$ days, which implies that $v_{\rm ej} \sim 5300\,{\rm km\,s^{-1}}$ at the semi-major axis of the ER (0\farcs{8}). 
Ejecta with higher velocities have already interacted with the ER, giving rise to hotspots in the ER, smaller spots outside the ER, as well as diffuse emission from the reverse shock. The morphology of the reverse shock was recently reconstructed in 3D based on James Webb Space Telescope (JWST) NIRSpec data, which showed that it extends from the inner edge of the ER to higher latitudes on both sides, forming a bubble-like structure \citep{larsson23}. The emission from the reverse shock at high latitudes is projected outside the ER in Fig.\,\ref{fig:general}. 

The very first hotspot in the ER appeared in 1995 in the northeast region, as a consequence of the SN blast wave crashing into the ER \citep{sonneborn98, lawrence00}. Additional spots appeared in the following years to reach a total of approximately 28\footnote{Out of these bright spots in the ER, the one in the southwest nearest $\text{PA}=230^\circ$ is a star projected on the ER.} covering the whole ER at 12\,980 days \citep{tegkelidis23}. These spots indicate the presence of gas clumps of higher density than the surrounding gas \citep{groningsson08}. The supernova ejecta, through their interaction with the ER, are now dissolving the hotspots \citep{fransson15}.

An increase in the optical fluxes in the ER was observed until about $8300$\,days after the explosion, followed by a decrease pretty much linear in time \citep{fransson15, larsson19a}. A similar behavior was observed at all wavelengths -- apart from radio and hard X-rays, which are still increasing in flux \citep{cendes18, alp21, maitra22}. New small spots outside the ER show that the blast wave has propagated beyond the ring \citep{larsson19a, arendt23}. Signs of shock interaction outside the ER are also clearly seen in JWST/NIRCam images \citep{arendt23}. The recent JWST/NIRSpec and MRS observations show that the inner ejecta have recently started to interact with the reverse shock at the inner edge of the ER in the southwest \citep{fransson24}. 

The regular monitoring by HST has provided a detailed record of the ejecta evolution. When the HST observations started in the early 1990s, the dominant radioactive isotope supplying energy input to the ejecta was $^{44}\text{Ti}$. This resulted in a slowly decreasing optical luminosity until about 5000\,days after the explosion.
The optical luminosity then started to increase in both the red and blue optical bands owing to the absorbed and thermalized X-rays from the ejecta interacting with the ring \citep{larsson11}, marking the transition to the supernova remnant phase. This transition is also marked by the change of morphology of the ejecta from a centrally dominated emission elongated in the northeast-southwest direction to a ``horseshoe-like'' shape. This horseshoe is irregularly elongated in the same direction but with a brightening of the outer limb, with the northern (respectively southern) lobe being predominantly blue- (respectively red-) shifted \citep[see Fig.\,\ref{fig:general}, but also][]{fransson13, larsson13}. The observed morphology of the ejecta is also expected to be affected by dust, which was observed by Herschel and ALMA \citep{matsuura15, cigan19}. 

A major new advancement in the study of SN\,1987A is the recent JWST observations that revealed the first unambiguous electromagnetic signal due to the compact object created in the explosion \citep{fransson24}. The observations show narrow, ionized emission lines from [Ar\,\textsc{ii}], [Ar\,\textsc{vi}], [S\,\textsc {iii}], and [S\,\textsc{iv}] at the center of the system, which can only be explained by ionization of the innermost, slow-moving ejecta by a compact central source. The exact nature of this source is yet to be determined, with the most likely scenarios being either a pulsar wind nebula or a cooling neutron star. Emission lines associated with the compact object are expected also at optical wavelengths and it is therefore of great interest to search for signs of this emission in the recent HST observations \citep[see][for previous upper limits]{alp18}. The probability of detecting the central source increases with time due to the expansion of the ejecta, which leads to gradually decreasing background emission as well as decreasing optical depth of the dust. We note that the central source may be revealed as a general diffuse brightening in the central region rather than as a point source if the effects of dust scattering are important.

As part of a long-term monitoring program, SN\,1987A was observed in September 2022 (days 12\,978 to 12\,980 after the explosion) by HST/WFC3 in nine filters, which cover the entire wavelength region between 2000 and 11\,000 \AA. It is the first time that HST imaging observations of SN\,1987A cover the full optical domain since 2009 (8329 days after the explosion). We make use of these and earlier archival observations to study the recent evolution of the broad- and narrow-band photometry and morphology of the ejecta between days 8329 and 12\,980. We also use the annual imaging in the F438W and F625W filters to provide light curves of the system, which adds four years of data compared to \citet{larsson19a}.
These different measurements reveal the asymmetric explosion geometry in increasingly great detail and provide information on the evolving energy sources. The latter are predicted to affect the temporal evolution in different ways. The energy input from the decay of $^{44}$Ti will lead to a gradual fading across the whole ejecta. Any significant contribution from the compact object would manifest as a brightening of the central region and a change of the relative brightness of different photometric bands. Finally, the energy input from X-rays from the ER is expected to give rise to changes in flux and morphology that reflect the X-ray evolution of the ER, as well as a general limb-brightened morphology (according to the model in \citealt{fransson13}).

\begin{figure}
\includegraphics[clip=true, trim=35 40 25 40,width=\linewidth]{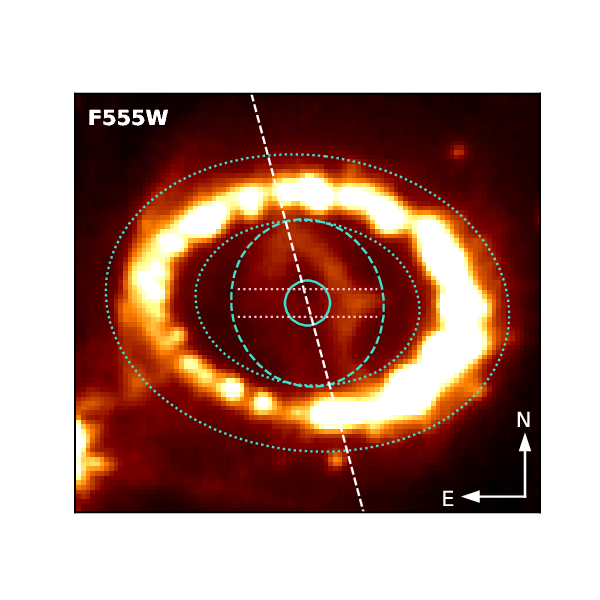}
\caption{HST/WFC3 2022 image of SN\,1987A in the F555W filter together with the regions adopted to compute the fluxes. The dotted cyan lines define the elliptical annulus around the ER, while the dashed and solid cyan lines define the elliptical region around the ejecta and the circular region around the center, respectively (see Sect.\,\ref{sect:SEDs}). The regions for the ejecta and the center were expanded over time to account for the expansion of the ejecta. The white dashed line represents the separation between eastern and western parts. The pink dotted lines represent the horizontal region adopted for Fig.\,\ref{fig:homologous_expansion}.} The image was scaled by an $\asinh$ function to highlight the weak emission outside the ER. The field of view is $2\fds50 \times 2\fds25$.\label{fig:regions}
\end{figure}

This paper is organized as follows. We describe the new 2022 HST observations and the data reduction process in Sect.\,\ref{sect:observations}. We discuss the lines contributing to the different filters in Sect.\,\ref{sect:lines}.  The change of morphology of SN\,1987A from days 8329 to 12\,980 is presented in Sect.\,\ref{sect:morphology}, while broad- and narrow-band photometry in three different regions, namely the ER, ejecta, and center (see the apertures delineated in Fig.\,\ref{fig:regions}) is presented in Sect.\,\ref{sect:SEDs}. Section\,\ref{sect:lightcurve} is devoted to the analysis of the light curve of SN\,1987A from day 8329 to day 12\,980 in the same three regions in the F438W and F625W filters. 
The results are discussed in Sect.\,\ref{sect:discussion} and we provide our conclusions in Sect.\,\ref{sect:conclusion}. 
 Throughout this paper, the distance to SN\,1987A is taken to be 49.59\,kpc \citep{pietrzynski19} and we refer to spectral lines by their vacuum wavelengths.

\section{Observations and data reduction\label{sect:observations}} 
\begin{table}[t]
\centering
\caption{Details of the HST/WFC3 observations. Columns~1 and 2 give the date of observation and the epoch measured in number of days since the explosion on 1987 February 23. Columns~3 and 4 give the filter and exposure time of the observation. The new WFC3 images are presented in the bottom section.}
\begin{tabular}{lllll}
\hline\hline
Date & Epoch & Filter & Exposure Time  \\
(YYYY month dd) & (days) & & (s) \\
\hline
2009 Dec 13 & 8329 & F225W & 800 \\
2009 Dec 13 & 8329 & F336W & 800 \\
2009 Dec 13 & 8329 & F438W & 800 \\
2009 Dec 12 & 8328 & F502N & 6200 \\
2009 Dec 13 & 8329 & F555W & 400 \\
2009 Dec 12 & 8328 & F625W & 3000 \\
2009 Dec 13 & 8329 & F657N & 1600 \\
2009 Dec 13 & 8329 & F814W & 400 \\
\hline
2011 Jan 05 & 8717 & F438W & 1400  \\
2011 Jan 05 & 8717 & F625W &  1140 \\
2013 Feb 06 & 9480 & F438W &  1200\\
2013 Feb 06 & 9480 & F625W &  1200\\
2014 Jun 15 & 9974 & F438W &  1200\\
2014 Jun 15 & 9974 & F625W &  1200 \\
2015 May 24 & 10\,317 & F438W &  1200\\
2015 May 24 &10\,317  & F625W &  1200\\
2016 Jun 08 & 10\,698 & F438W &  600\\
2016 Jun 08 & 10\,698 & F625W &  600\\
2017 Aug 03 & 11\,119 & F438W &  1400\\
2017 Aug 03 & 11\,119 & F625W &  1200\\
2018 Jul 08 & 11\,458 & F438W &  1200\\
2018 Jul 08 & 11\,458 & F625W &  1200\\
2019 Jul 22 & 11\,837 & F438W &  1200\\
2019 Jul 22 & 11\,837 & F625W &  1200\\
2020 Aug 06 & 12\,218 & F438W &  1160\\
2020 Aug 06 & 12\,218 & F625W &  1160\\
2021 Aug 21 & 12\,598 & F438W &  1200\\
2021 Aug 21 & 12\,598 & F625W &  1080\\
\hline
2022 Sep 06 & 12\,979 & F275W & 2800 \\
2022 Sep 06 & 12\,979 & F280N & 5600 \\
2022 Sep 06 & 12\,979 & F336W & 2600 \\
2022 Sep 05 & 12\,978 & F438W & 1200 \\
2022 Sep 07 & 12\,980 & F502N & 5600 \\
2022 Sep 05 & 12\,978 & F555W & 1080 \\
2022 Sep 05 & 12\,978 & F625W & 1080 \\
2022 Sep 06 & 12\,979 & F657N & 2600 \\
2022 Sep 05 & 12\,978 & F814W & 1080 \\
\hline
\end{tabular}
\label{table:obs}
\end{table} 

The HST/WFC3 observations analyzed in this paper consist of nine new images taken in September 2022, eight archival images taken in December 2009 (seven out of these used filters in common with the 2022 images), and 20 archival images taken every year between 2011 and 2021, except in 2012, in the two wide-band filters F438W and F625W (details provided in Table\,\ref{table:obs}). In Table\,\ref{table:filters}, we provide the passband rectangular widths and the pivot wavelengths of the ten different filters. The HST observations between days 8329 and 11\,458 were analyzed in a series of papers to which we refer for more detailed information \citep{larsson11, larsson13, larsson19a, fransson15, france15, alp18}.

\begin{table}[t]
\centering 
\caption{Properties of the HST/WFC3 filters. Columns 1, 2, and 3 give the name, the passband rectangular width, and the pivot wavelength of the filter, respectively.}
\begin{tabular}{lll}
\hline\hline
Filter & Width$^\text{a}$ (\AA) & Pivot wavelength$^\text{b}$ (\AA) \\
\hline
F225W & 467 & 2372\\
F275W & 405 & 2710 \\ 
F280N & 42 & 2833 \\
F336W & 512 & 3355 \\ 
F438W & 615 & 4326 \\ 
F502N & 65 & 5010 \\ 
F555W & 1565 & 5308 \\ 
F625W & 1465 & 6243 \\
F657N & 121 & 6567 \\ 
F814W & 1565 & 8039 \\
\hline
\end{tabular}
\begin{tablenotes}
\item $^\text{a}$Defined as the equivalent width divided by the maximum throughput within the filter bandpass.
\item $^\text{b}$Measure of the effective wavelength of a filter, calculated based on the integrated system throughput.
\end{tablenotes}
\label{table:filters}
\end{table}

The dithered exposures were combined using \texttt{DrizzlePac} \citep{ferland13} adopting a pixel scale of $0\fds025$. Cosmic rays were removed and distortion corrections were applied as part of the drizzling process \citep{gonzaga12}. The HST astrometry was corrected by aligning field stars with GAIA as in \citet{larsson19a}. The images were aligned to the same pixel frame using \texttt{astropy reproject} with bilinear interpolation.

\section{Line contributions to the different filters\label{sect:lines}}
\begin{figure*}
\centering
\includegraphics[clip=true,trim=0 0 0 0,width=0.8\linewidth]{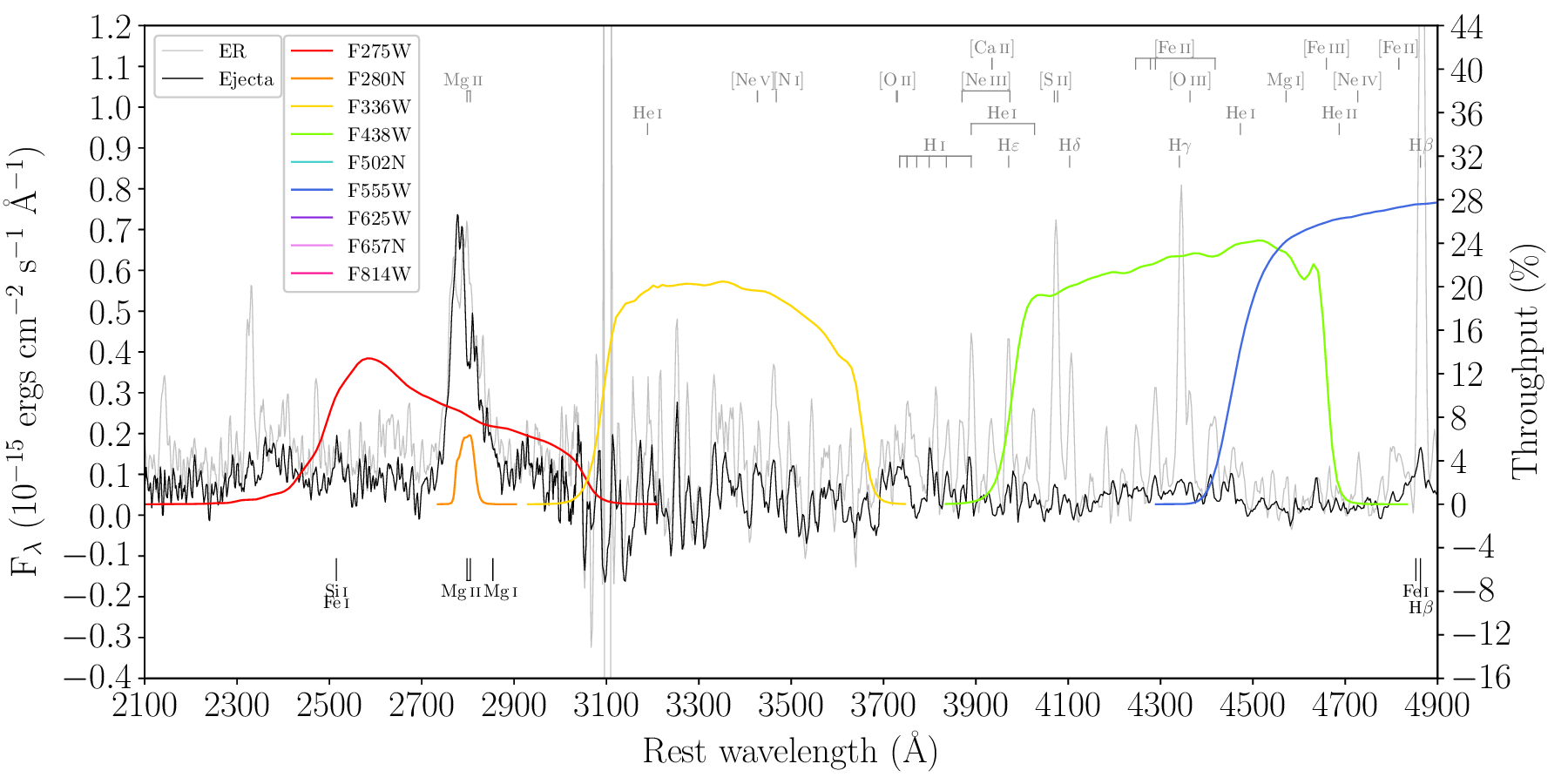}
\includegraphics[clip=true,trim=0 0 0 0,width=0.8\linewidth]{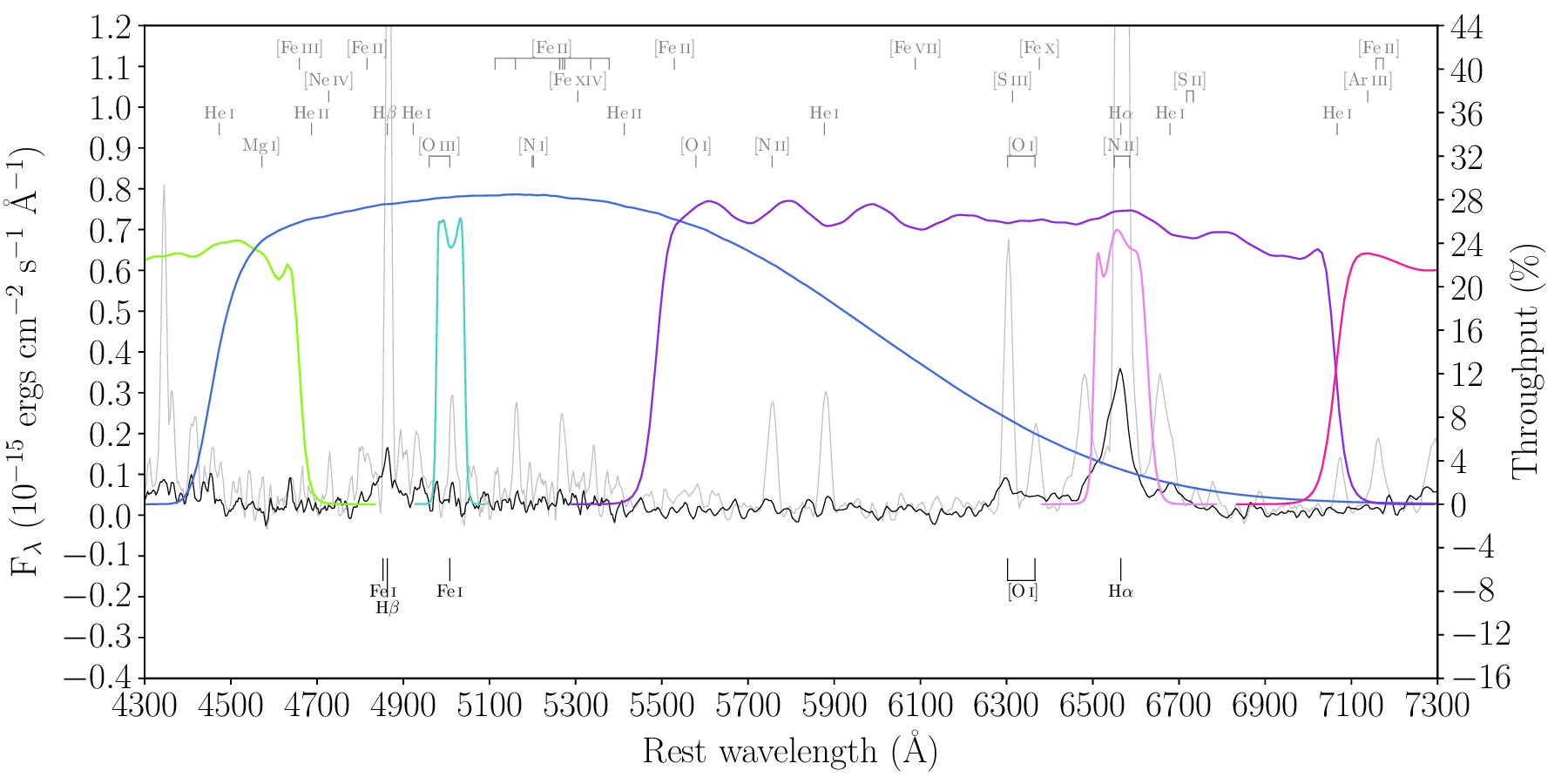}
\includegraphics[clip=true,trim=0 0 0 0,width=0.8\linewidth]{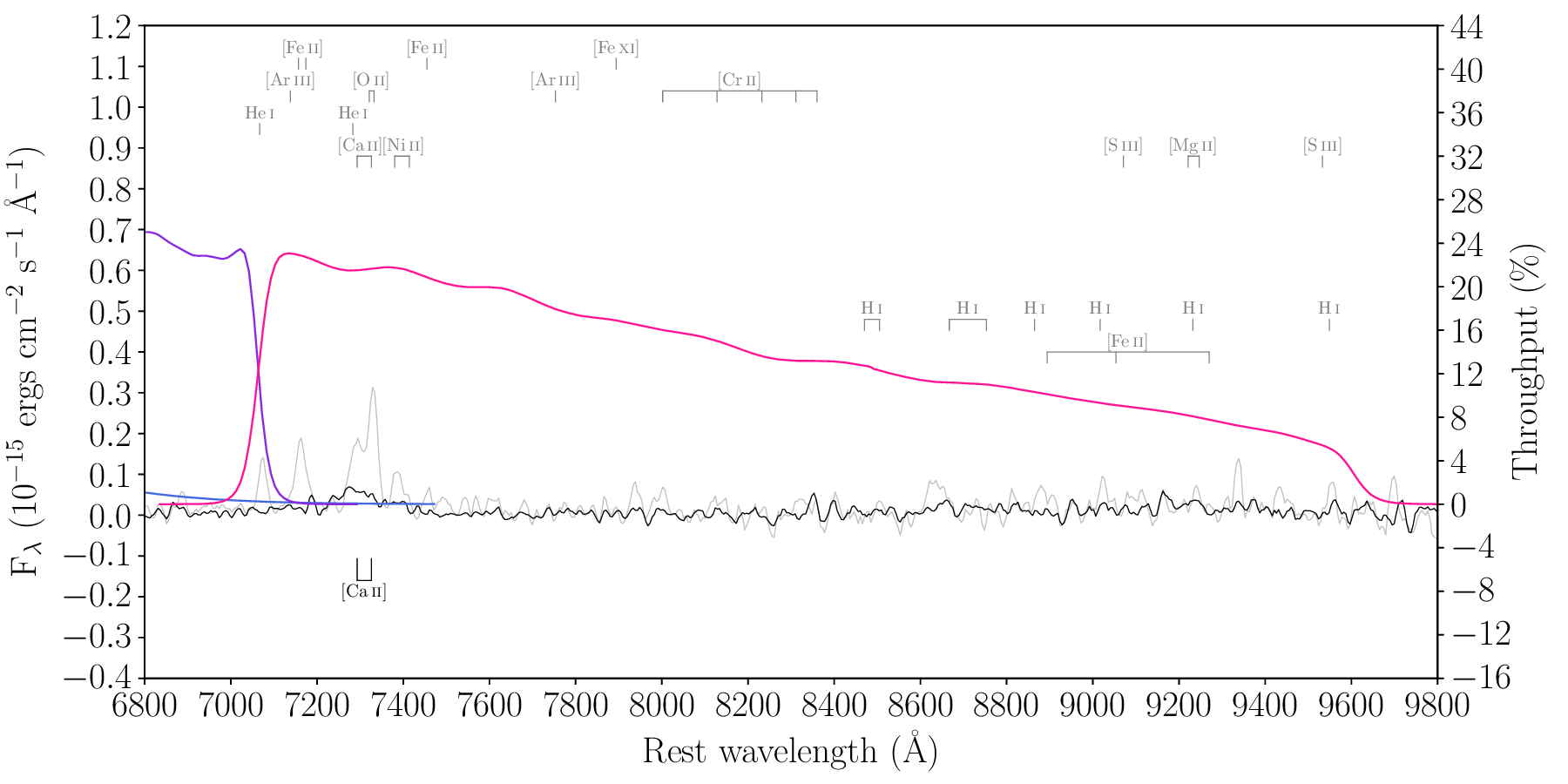}
\caption{STIS spectra from 2017 \citep[described in][]{kangas22} of the ER (gray) and the ejecta (black), and HST/WFC3/UVIS2 filters' response functions (in colors). The spectra were obtained from two adjacent $0\fds2$-wide slits oriented in the north-south direction, covering the central part of the remnant. They were smoothed with the Savitzky-Golay algorithm for visual clarity. Lines contributing to the ejecta and the ER are identified in black and gray, below and above the spectra, respectively. \label{fig:spectra}}
\end{figure*}

To identify which lines contribute to the different filters, we present in Fig.\,\ref{fig:spectra}, together with the filters' response functions, the Space Telescope Imaging Spectrograph (STIS) spectra from 2017 of the ER and the ejecta discussed in \citet{kangas22}. The STIS spectra were obtained from two adjacent $0\fds2$-wide slits oriented in the north-south direction. The extraction regions cover most of the inner ejecta, as well as the north and south parts of the ER. We added the fluxes in the two slits to form the spectra shown in Fig.\,\ref{fig:spectra}. We used the STIS spectra here despite their relatively low signal to noise ratio (S/N) because they are the most recent spectra covering the whole wavelength region considered here. Higher S/N optical spectra of the ejecta taken at earlier times (1995) were modeled in \citet{jerkstrand11}. We refer to this paper for the identification of lines that are close to noise level in the 2017 spectra. 
The optical spectra of the ER and the ejecta are both dominated by line emission; the continuum only contributes a small amount to the total flux in any spectral region or filter \citep[see][]{groningsson08, jerkstrand11}.

The narrow-band filters are mostly dominated by a single line: Mg\,\textsc{ii}~$\lambda\lambda\,2795, 2802$ in F280N, Fe\,\textsc{i}~$\lambda\,5007$ (ejecta) or [O\,\textsc{iii}]~$\lambda\,5008$  (ER) in F502N, and H$\alpha$ in F657N. The narrow lines from the ER are fully contained by the filters (except for F280N), while the broad ejecta lines are not. The velocity ranges for the main emission lines in the F280N, F502N, and F657N filters are given in Table\,\ref{table:velocity}.

\begin{table*}
\centering
\caption{Minimum and maximum velocities for the main emission lines of the ejecta in the narrow-band filters at 50\% and 10\% of the peak transmission.}
\begin{tabular}{lllllll}
\hline\hline
Filter & Ion & Reference wavelength & $v_{\text{min,10\%}}$ & $v_{\text{min,50\%}}$ & $v_{\text{max,50\%}}$ & $v_{\text{max,10\%}}$ \\
& & (\AA) & (km\,s$^{-1}$) & (km\,s$^{-1}$) & (km\,s$^{-1}$) & (km\,s$^{-1}$) \\
\hline
F280N & Mg\,\textsc{ii} &2799 & -3443 & -2854 & +1643 & +2767 \\
F502N & [O\,\textsc{iii}] & 5008 & -2922 & -2097 & +1863 & +2270 \\ 
F657N & H$\alpha$ & 6565 & -3518 & -3061 & +2681 & +3685 \\
\hline
\end{tabular}
\label{table:velocity}
\end{table*}

\begin{figure*}
\centering
\includegraphics[clip=true,trim=0 60 0 0,width=\linewidth]{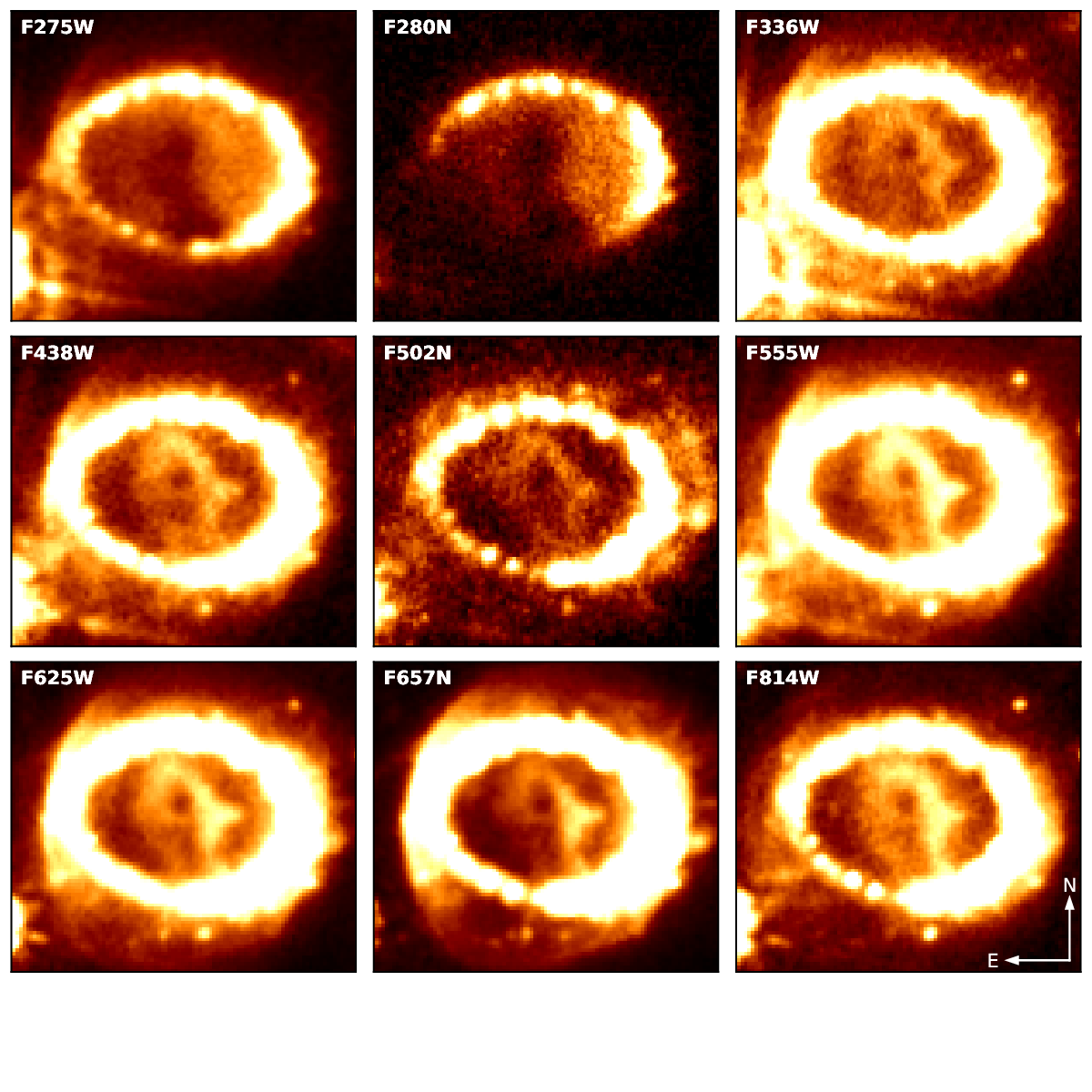}
\caption{HST/WFC3 images of SN\,1987A taken 12\,980 days after the explosion in nine different filters. The images were scaled by an $\asinh$ function and the color scales were chosen differently for each filter to highlight the weak emission in the ejecta. The field of view for each  image is $2\fds50 \times 2\fds25$.\label{fig:ejecta_all_2022}}
\end{figure*}

The broad-band filters cover many lines that might differ between the ejecta and the ER. Regarding the ejecta, scattering by the Mg\,\textsc{ii}~$\lambda\lambda\,2795, 2802$ lines dominate in the F275W filter. The formation of these fluorescent lines is partly powered by Ly$\alpha$, which is absorbed in the Mg\,\textsc{ii}~$\lambda\,1240$ line, which then in turn produces the Mg\,\textsc{ii}~$\lambda\lambda\,2795, 2802$ lines \citep[see discussion in Sect.\,\ref{subsect:lineexcitation}, but also][]{kangas22}. We here note that the F225W filter used in the 2009 observations (presented later in Fig.\,\ref{fig:ejecta_all_2009}) mainly contains Fe\,\textsc{i}, Fe\,\textsc{ii}, and Si\,\textsc{i} lines \citep{jerkstrand11}, while the Mg\,\textsc{ii} lines only slightly contribute to this filter. No single line dominates in the F336W filter, and no clear line identification can be made in the range between 3000 and 3500\,\AA\ as asserted by \citet{jerkstrand11}. Instead, the F336W filter covers a blend of scattering and fluorescence lines, mostly from Fe: The emission is a mix of mainly H\,\textsc{i} and Fe\,\textsc{i-ii} lines, but also scattering from resonance lines of trace elements in the H envelope. The range covered by the F438W filter is a blend of Fe\,\textsc{i}, Ca\,\textsc{i}, H\,\textsc{i}, notably the  Ca\,\textsc{i}~$\lambda\,4226$ emission line, complemented with emission and scattering from the Fe/He zone ([Fe\,\textsc{ii}]~$\lambda\lambda\,4223, 4339, 4453$), and H$\gamma$~$\lambda\,4343$. The F555W filter is largely dominated by the H$\beta$ line, though Fe\,\textsc{i} also contributes to the feature at 4850\,\AA. 
The H$\alpha$ line dominates in the F625W filter, and also contributes to the F555W filter. The [O\,\textsc{i}]~$\lambda\lambda\,6300, 6364$ lines contribute to both the F555W and F625W filters, though their fluxes are contaminated by Fe\,\textsc{i} lines in this wavelength range. The F814W filter is dominated by the [Ca\,\textsc{ii}]~$\lambda\lambda\,7293, 7326$ lines.

The ER has many strong lines that also appear in the ejecta. Additional lines contribute to the different filters (see Fig.\,\ref{fig:spectra}), and we refer the reader to \citet{groningsson08} for a thorough identification of the emission lines from the ER. 

\section{Morphology\label{sect:morphology}}
The HST/WFC3 observations of SN\,1987A at day 12\,980 after the explosion are presented in Figs\,\ref{fig:ejecta_all_2022} and \ref{fig:ER_all_2022} (optimized to see the ejecta and ER, respectively).
The morphology of the ER at epoch 12\,980 days is very similar in all filters: brightest in the west and faintest in the southeast. Part of the ER is missing in the F280N filter due to the narrow filter not covering all Doppler shifts (see Sect.\,\ref{sect:lines}). Emission is observed outside the ER, including the reverse shock (especially in filters with strong H lines) and an outer portion on the west side in the F502N filter. These emission components will be studied in a separate paper. 

\subsection{Morphologies of the ejecta at day 12\,980 after the explosion\label{subsect:morphologies}}
Overall, the ejecta appearance is very similar in all filters, with the exception of the Mg\,\textsc{ii}-dominated ones (F275W and F280N). 
The ejecta show a north-south elongation where it is brightest. The western part of the ejecta appears brighter than the eastern part, while a small dimmer hole, a region actually called ``the hole", located slightly north to the center of the ejecta, appears. These observations are very similar to the HST observations taken during the last few years (see Fig.\,\ref{fig:center_f438w_2009_to_2022} discussed in Sect.\,\ref{sect:lightcurve} and Appendix\,\ref{appendix:additionalfigures}).

\begin{figure}
\centering
\includegraphics[clip=true,trim=0 30 0 30,width=1\linewidth]{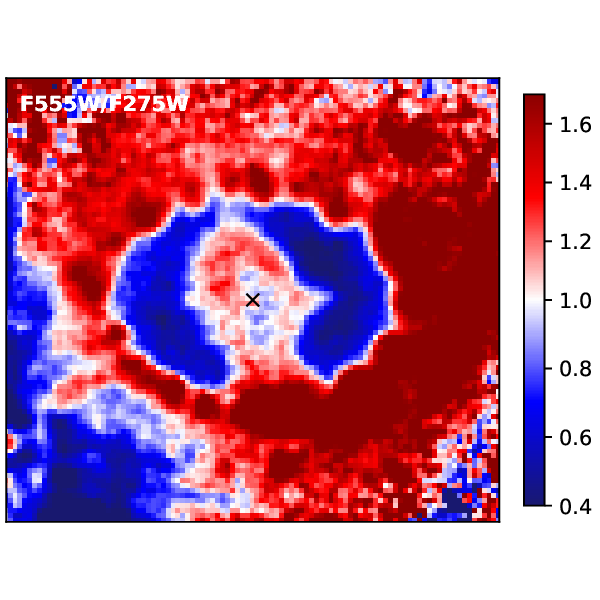}
\includegraphics[clip=true,trim=0 30 0 30,width=1\linewidth]{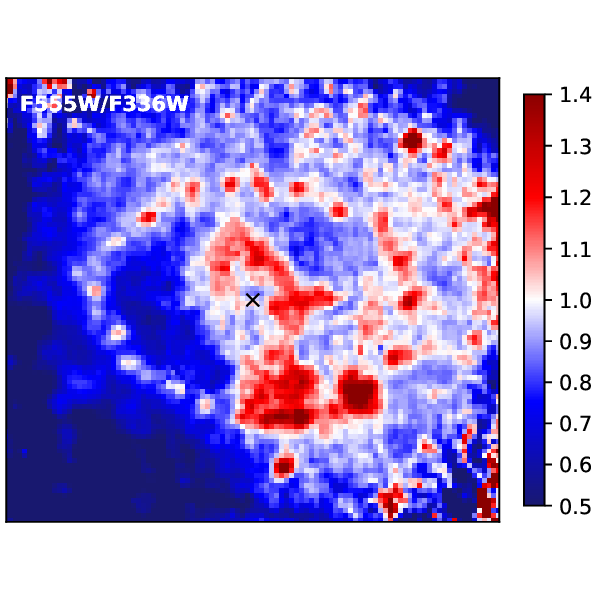}
\includegraphics[clip=true,trim=0 30 0 30,width=1\linewidth]{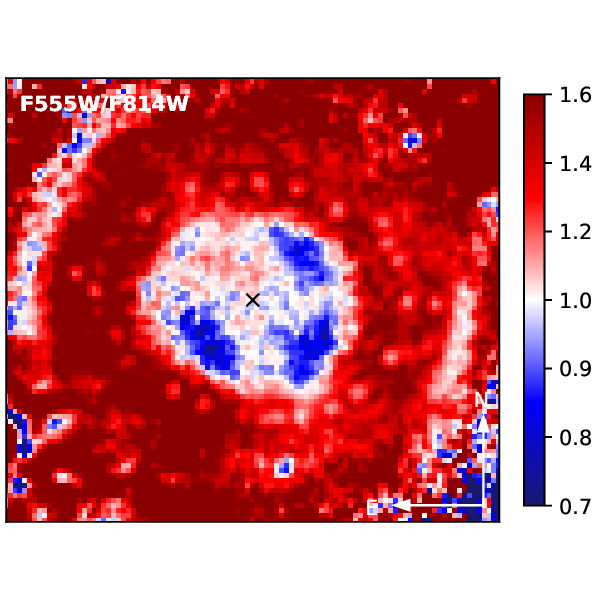}
\caption{Ratio of HST/WFC3 images of SN\,1987A taken 12\,980 days after the explosion scaled by the ratio of count rates in the center region (note the two linear scales for values $\ge1$ and $\le1$, respectively). The highest resolution images were degraded in resolution to match the lowest resolution images (see main text for details). The field of view for each image is $2\fds50 \times 2\fds25$. The black cross indicates the geometric center of the ejecta. \label{fig:ratio_images}}
\end{figure}

\begin{figure*}
\centering
\includegraphics[clip=true,trim=0 60 0 0,width=\linewidth]{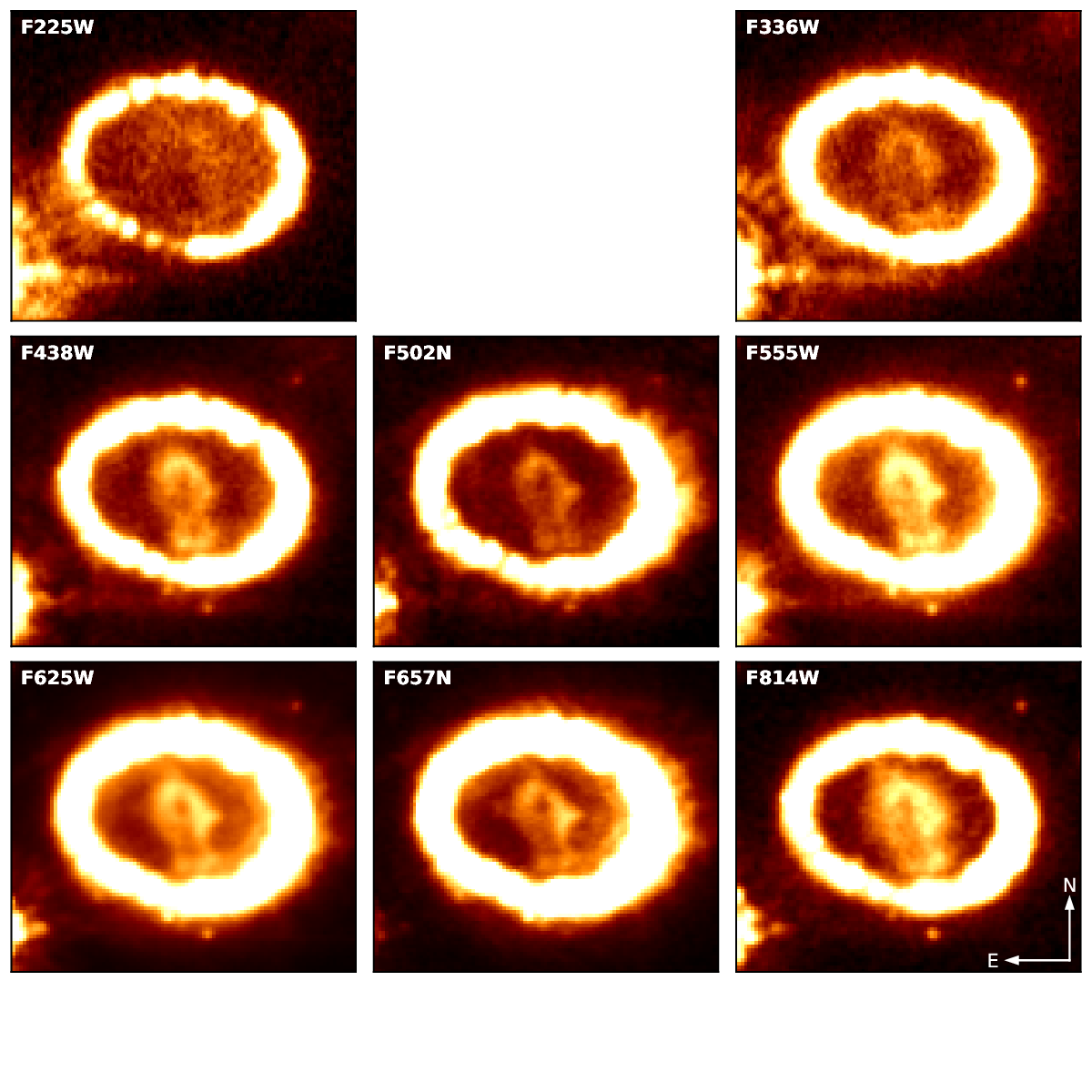}
\caption{HST/WFC3 images of SN\,1987A taken 8329 days after the explosion in eight different filters. The images were scaled by an $\asinh$ function and the color scales were chosen differently for each filter to highlight the weak emission in the ejecta. The scales are twice as large as those of the corresponding 2022 images (see Fig.\,\ref{fig:ejecta_all_2022}). The field of view for each  image is $2\fds50 \times 2\fds25$.\label{fig:ejecta_all_2009}}
\end{figure*}

Given that Balmer hydrogen lines mostly dominate in the F555W, F625W, and F657N filters, it is not surprising that the observations taken in these filters are very similar in terms of morphology, though the F657N image looks a little different from the F625W due to the limited filter width (some of the ejecta are Doppler shifted outside, see Sect.\,\ref{sect:lines}). A similar overall morphology is observed in the F336W, F438W, and F814W filters, though it is notable that the S/N in the ejecta is low in the former two filters (see Sect.\,\ref{sect:lines}). The ejecta morphology in the F502N filter is strongly affected by the narrow filter width which only covers the Fe\,\textsc{i} line.
Finally, the morphology of the ejecta in the F275W and F280N filters differs from the other filters because of Mg\,\textsc{ii} resonance line scattering \citep[discussed in][based on the STIS spectra]{kangas22}, which can occur in both the metal core and the H and He envelopes of the pre-explosion progenitor star. 

To better identify the differences in morphology between the filters, we investigated the ratio between pairs of images. We accounted for the small wavelength dependence of the spatial resolution (the FWHM varies between $0\fds086$ and $0\fds095$ depending on the filter) by degrading the resolution of the high-resolution image to match the corresponding low-resolution image. We used the \texttt{photutils PSF matching} package (with a Hanning window) in \texttt{python} and adopted a well-isolated star in the field of view as our reference for the point spread function (PSF). These matched kernels were convolved with the \texttt{python} package \texttt{scipy ndimage} with a convolution mode “nearest”. We then scaled the image ratios by the corresponding count rates in the center region to better highlight the differences in morphology between the different filters. The main results are presented in Fig.\,\ref{fig:ratio_images} and discussed hereafter.  We note that this PSF matching is not perfect: Some residuals are left in the images, mainly at the locations of the hotspots (see the relative brightness difference between the ER spots' interior and rim), while the ejecta seem to be free of these artefacts.

Compared to the filter F555W, the region around the inner ejecta but inside the ER is proportionally much brighter in the F275W filter (see upper panel of Fig.\,\ref{fig:ratio_images}), especially in the western part. The central region of the ejecta is also proportionally slightly brighter than the rest of the ejecta in the F275W filter compared to the F555W filter. The spots, especially the south-western ones, in the F275W are proportionally fainter.  
A slight relative difference in brightness is observed between the filters F336W and F555W: The region east (respectively west) of the ejecta but inside the ER is proportionally slightly brighter (respectively fainter) in the F336W filter, highlighting the shape of the ejecta (see middle panel in Fig.\,\ref{fig:ratio_images}).  
The ejecta are proportionally brighter than the ER in the F814W filter compared to the F555W filter (see lower panel in Fig.\,\ref{fig:ratio_images}). From the overall similarity, we infer the ejecta are chemically well-mixed on large scales.

\subsection{Evolution of morphologies of the ejecta between days 8329 and 12\,980 after the explosion\label{subsect:morphologies_evolution}}
\begin{figure*}
\centering
\includegraphics[clip=true,trim=40 5 5 5,width=1\linewidth]{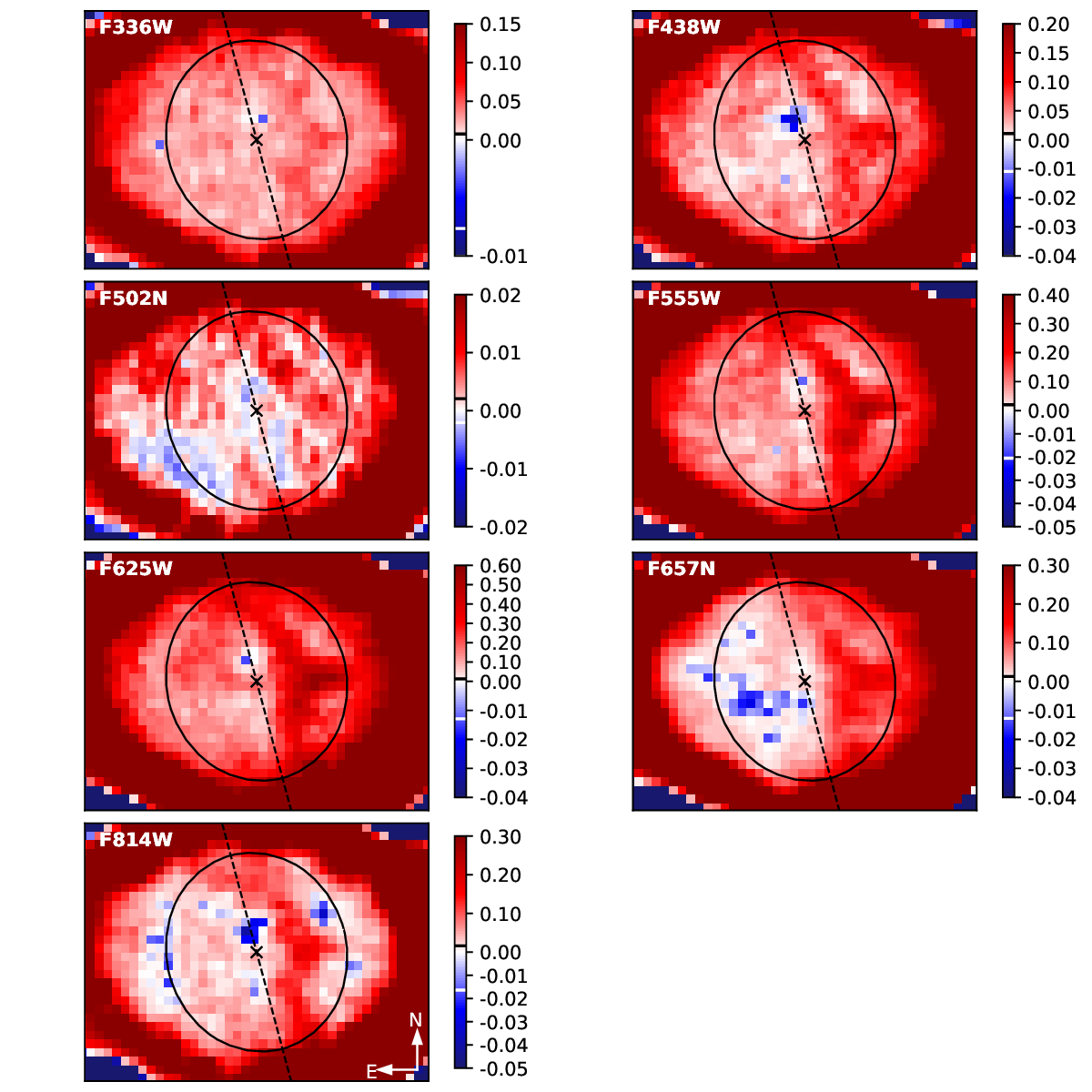}
\caption{Differences of SN\,1987A HST/WFC3 images between epochs 12\,980 and 8329 days after the explosion in the seven common filters. The observations at day 12\,980 were reduced in size and ``re-observed" at the epoch of the  observations at day 8329 assuming the ejecta were expanding freely, before subtraction in order to highlight the change of morphology in the ejecta. To account for the changing ER contribution to the ejecta, an offset of $-0.0063$, $-0.0083$, $-0.0016$, $-0.0196$, $-0.0319$, $-0.0230$, and $-0.0097$, respectively (corresponding to the mean flux of the ER inside the ejecta as computed from the synthetic models for the ER, see Appendix\,\ref{appendix:ER_synthetic}), was applied to the images. The aperture defining the ejecta region is overplotted on the images as a black ellipse. The dashed line represents the separation between the eastern and western parts. The field of view for each of the images is $0\fds500 \times 0\fds375$. Values are given in units of electrons per seconds (note the two different, but linear, intensity scales for positive and negative values, with 0.0 as the crossover point). The black and white lines in the colorbars indicate the $3\sigma$ levels above and below which differences are significant, calculated from the number of counts in individual pixels in the ejecta region.}\label{fig:2022_shrinked_minus_2009}
\end{figure*}

The HST/WFC3 observations of SN\,1987A at day 8329 after the explosion  are presented in Figs\,\ref{fig:ejecta_all_2009} and \ref{fig:ER_all_2009} (optimized to see the ejecta and ER, respectively). We refer to \citet{larsson13} for a thorough analysis of these observations. We compare the morphologies of the HST observations taken at day 12\,980 to those taken at day 8329 in the seven common filters, namely F336W, F438W, F502N, F555W, F625W, F657N, and F814W, assuming the ejecta were expanding freely at constant velocity following Eq.\,\eqref{eq:ejectavel}. To this aim, the observations at day 12\,980 were resampled by a factor of 1.56, corresponding to the ratio between the number of days after explosion of the two epochs of observations. In practice, we drizzled the dithered exposures assuming a pixel size 1.56 times larger than the pixel size of $0\fds025$ adopted in the rest of the paper (see Sect.\,\ref{sect:observations}), all other parameters identical. As such, it is as if the epoch 12\,980 observations were run backwards in time and ``re-observed" at epoch 8329. In this process, the resulting images have the same number of pixels in the region covering SN\,1987A.

We then subtracted the observations at day 8329 from the corresponding reduced observations at day 12\,980, the results of which are presented in Fig.\,\ref{fig:2022_shrinked_minus_2009}.  
We multiplied the images by a scale factor to account for the small (maximum 2\%) decrease in instrumental sensitivity between the observations. Other changes related to the performance of WFC3 are handled in the data reduction process.
We here note that only the region of the ejecta is relevant in this context, as the ER in the more recent observations was artificially reduced in size in the process and therefore does not have any physical meaning.\footnote{We could rather expand the 2009 images by the same 1.56 factor. However, in that case, rather than reducing the resolution of the ejecta in the 2022 images to the 2009 level, we would artificially increase the resolution of the 2009 images to the 2022 level.}
The images indicate dimming in ``the hole" just above the central region in all filters, though it is significant at the $3\sigma$ level only in  filters F438W, F502N, and F814W. In all filters, we observe a clear brightening of the west part of the inner ejecta compared to the rest of the inner ejecta.
These two features point toward the presence of an external source of energy powering the ejecta in addition to the radioactive decay of $^{44}\text{Ti}$, as no such structure would clearly stand out if radioactivity was the main source of energy powering the freely expanding ejecta (see discussion in Sect.\,\ref{subsect:energysources}).

\section{Broad- and narrow-band photometry\label{sect:SEDs}}
In this section we analyze the broad- and narrow-band photometry in the three different regions presented in Fig.\,\ref{fig:regions}, namely the ER, the ejecta, and the center in the different filters in the observations taken at epoch 12\,980 and compare them to the observations at epoch 8329. All regions are centered on the supernova geometric center \citep[$\alpha=05^\text{h} 35^\text{m} 27\fdsec 9875$, $\delta=-69^\circ 16' 11\fds107$,][]{alp18}. 

The flux in the ER was computed assuming the same elliptical annulus for all observations, that is to say, assuming the expansion velocity of the ER is negligible\footnote{The expansion velocity of the ER was measured in \citet{larsson19a}: The semi-major axis expands at $680\pm50$\,km\,s$^{-1}$, corresponding to 3\,mas\,yr$^{-1}$, that is to say, 0.12 pixels per year. Over the 12.73 years from day 8329 to day 12\,980, this would total to 1.5 pixels of expansion.  This number of pixels is  negligible compared to our adopted aperture which is large enough to accomodate a 1.5 pixels expansion from day 8329 through all epochs up to day 12\,980.}. The region was chosen to be sufficiently large to encompass the bright emission seen in all observations, and is represented by the dotted cyan lines in Fig.\,\ref{fig:regions}. The annulus has an aspect ratio of 0.73, a position angle of 173$^\circ$, and its internal and external major axes equal to $1\fds21$ and $2\fds18$, respectively.

\begin{figure*}
\centering
\includegraphics[clip=true,trim=0 0 0 0,width=1\linewidth]{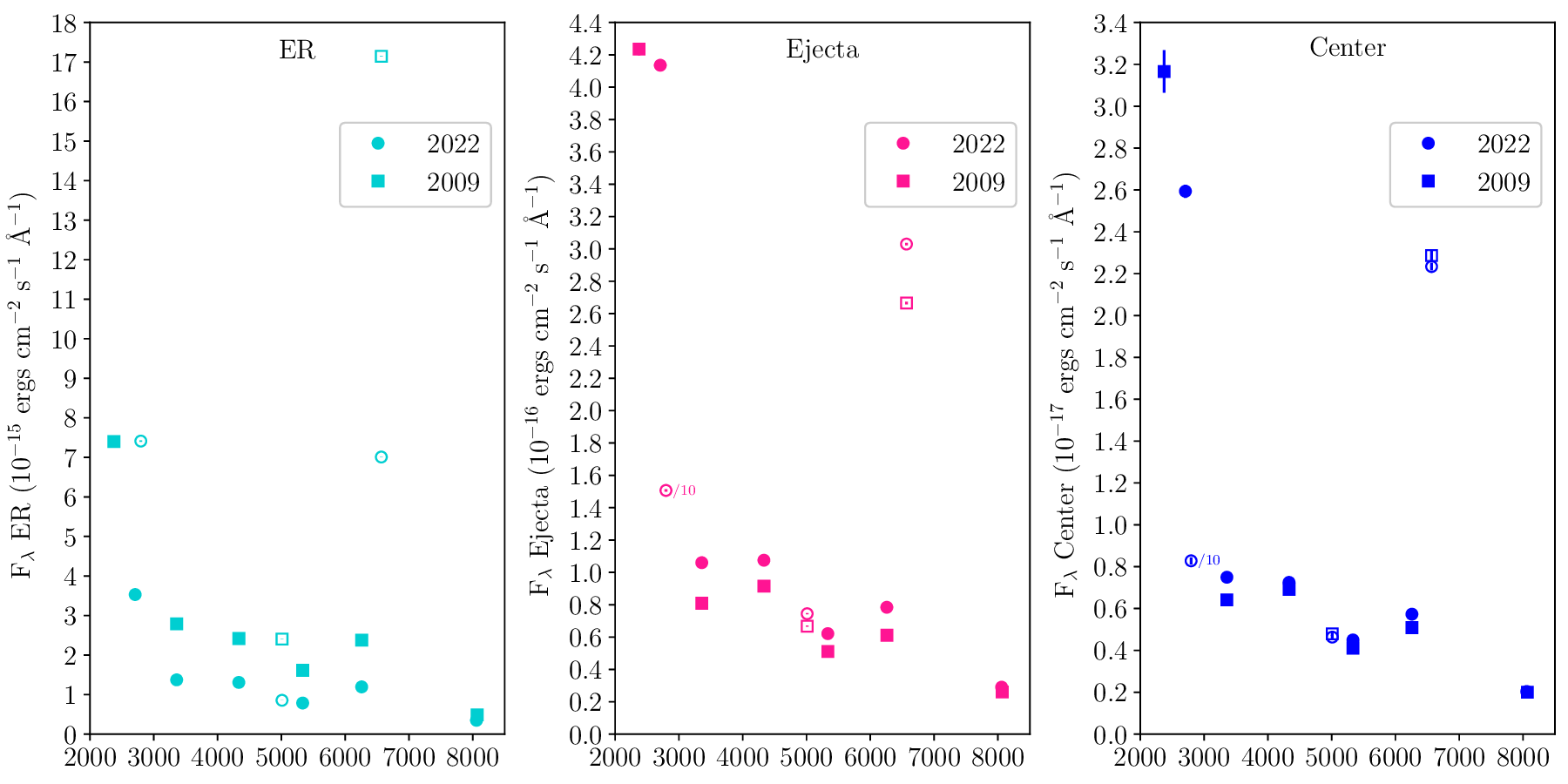}
\includegraphics[clip=true,trim=0 0 0 0,width=1\linewidth]{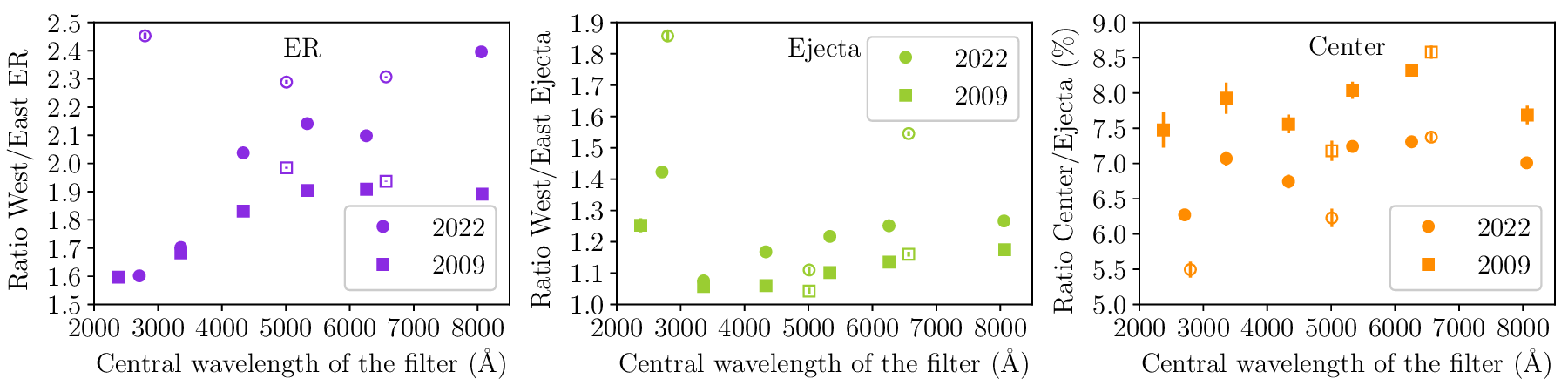}
\caption{Broad- and narrow-band photometry of the ER (\textit{top left panel}), ejecta (\textit{top middle panel}), and center (\textit{top right panel}). Corresponding ratios between western and eastern parts in the ER (\textit{bottom left panel}) and ejecta (\textit{bottom middle panel}), and ratio between the center and full ejecta fluxes (\textit{bottom right panel}). The observations at epochs 12\,980 and 8329 (labeled 2022 and 2009, respectively) are marked with circles and squares, respectively, while wide- and narrow-band filters are shown with filled and open symbols, respectively. We note that the error bars are smaller than the symbols in most cases. Since the shortest wavelength filters are different in the two epochs, they are plotted side by side and not included in Fig.\,\ref{fig:ratio_SEDs_2009_and_2022}.
\label{fig:SEDs_2009_and_2022}}
\end{figure*}

\begin{figure}
\centering
\includegraphics[clip=true,trim=0 0 0 0,width=1\linewidth]{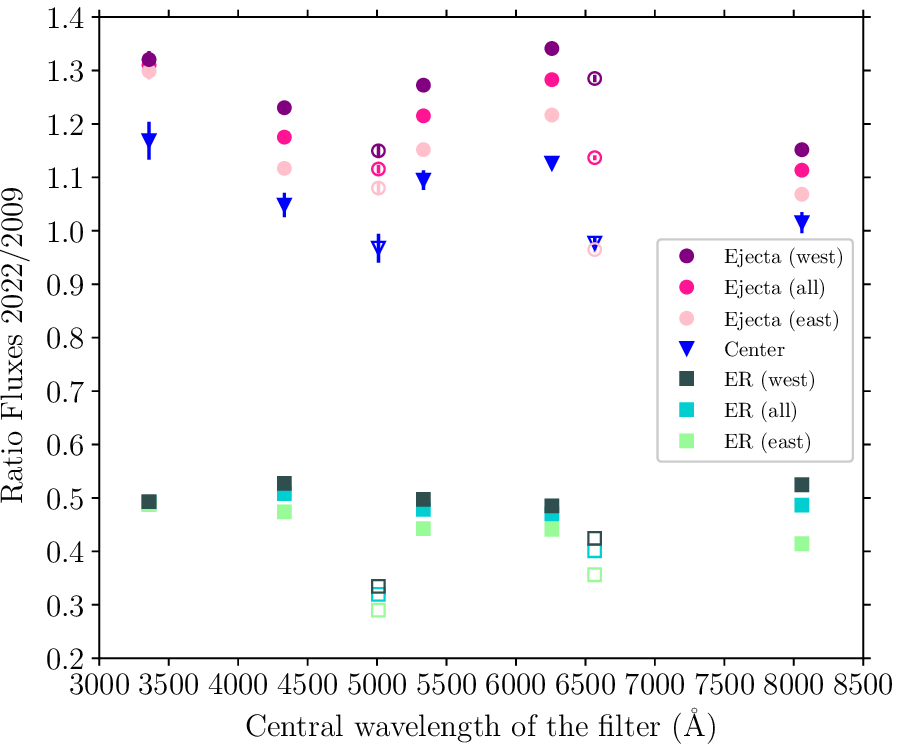}
\caption{Ratio between the fluxes at epochs 12\,980 and 8329 for the seven common filters in the ER (full region and western and eastern parts), ejecta (full region and western and eastern parts), and center. Wide- and narrow-band filters are shown with filled and open symbols, respectively.   \label{fig:ratio_SEDs_2009_and_2022}}
\end{figure}

\begin{figure*}
\centering
\includegraphics[clip=true,trim=0 0 0 190,width=1\linewidth]{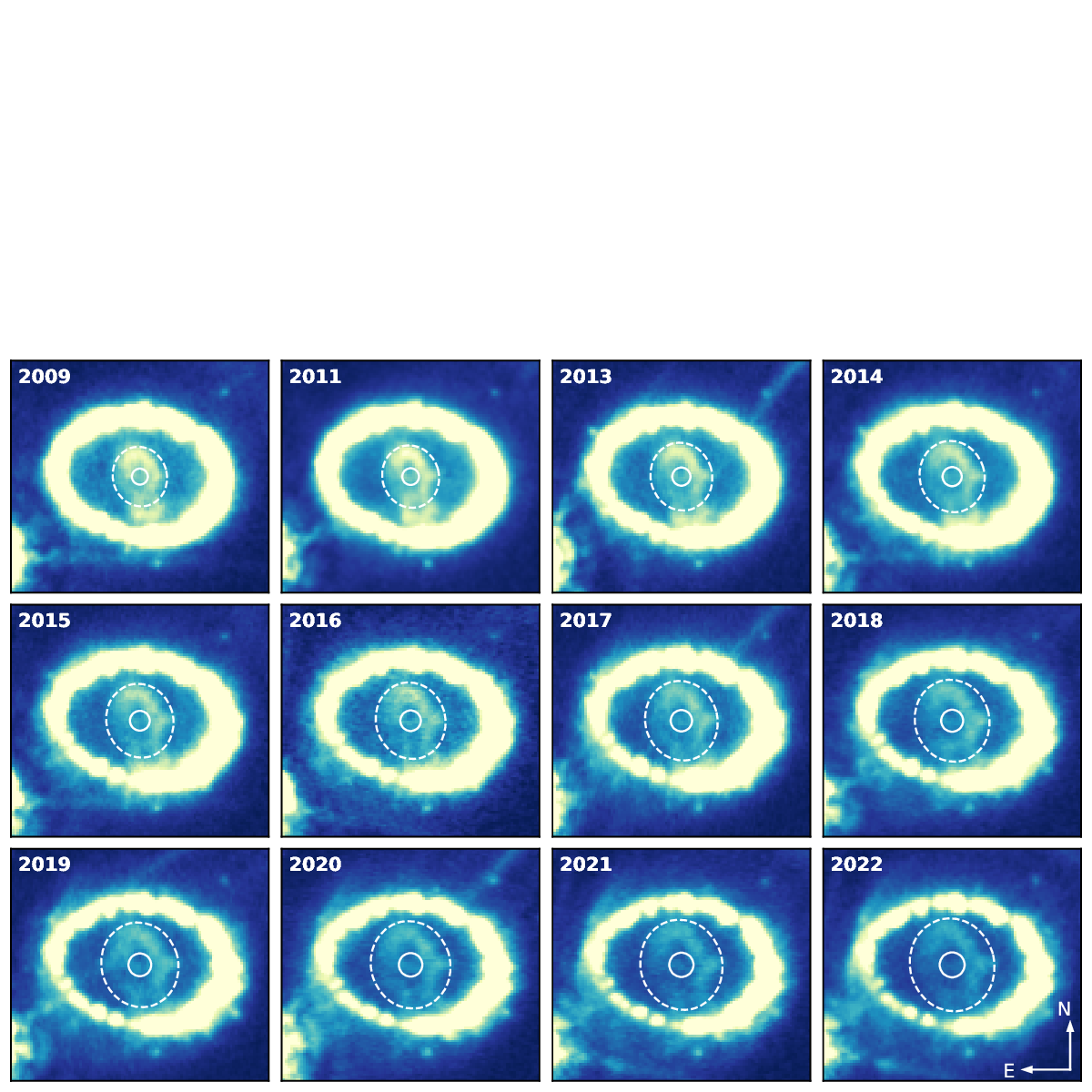}
\caption{HST/WFC3 images showing the evolution of SN\,1987A in the F438W filter between epochs 8329 and 12\,980 (labeled by the year of observation). The emission in the lower left corner is due to Star\,3 (see Fig.\,\ref{fig:general}). The images were scaled by an $\asinh$ function to highlight the weak emission in the ejecta. The field of view for each image is $2\fds50 \times 2\fds25$. The white circle and dashed white ellipse overplotted on each image represent the center and ejecta regions, respectively, adopted to compute the fluxes. These regions grow in size from 2009 to 2022, using Eq.\,\eqref{eq:ejectavel} to define a co-moving volume that tracks the same expanding ejecta with time.} \label{fig:center_f438w_2009_to_2022}
\end{figure*}

The flux in the ejecta was computed in an elliptical region defined by a major axis corresponding to an ejecta speed of 3000\,km\,s$^{-1}$, an aspect ratio of 0.90 and a positions angle of 105$^\circ$ \citep{larsson11}, as depicted by the dashed cyan line in Fig.\,\ref{fig:regions}. The fluxes reported for the ejecta region throughout the paper include all of the flux inside the elliptic aperture schematized in Fig.\,\ref{fig:regions}, and thus include the flux in the center too. This size of the ejecta region allows us to capture most of the bright inner ejecta, while avoiding the ER. The flux in the center was computed in a circular region defined by a radius corresponding to an ejecta speed of 800\,km\,s$^{-1}$, as depicted by the cyan solid line in Fig.\,\ref{fig:regions}. The motivation for looking at this specific region is that it should cover the compact object, considering a plausible range of kick velocities \citep{alp18, fransson24}. We here point out that defining the regions in terms of expansion velocity means that they are smaller (in arcsec) in the observations taken at epoch 8329 according to Eq.\,\eqref{eq:ejectavel}, as notably shown in Fig\,\ref{fig:center_f438w_2009_to_2022}. 

We summed the count rates over the pixels inside the given regions, accounting for fractional pixels. The $1\sigma$ statistical uncertainties are non-negligible (and bigger than the plot symbols) only in the center region. The count rates were converted into fluxes using the inverse sensitivity of the filters. Finally, the fluxes were de-reddened adopting a Milky Way extinction curve of \citet{cardelli89} with a color excess $E(B-V) = 0.19$\,mag and a reddening factor in the $V$-filter $R_V = 3.1$, as suggested by \citet{france11}. 

\begin{table*}
\centering
\caption{De-reddened flux measurements in the ER, ejecta, and center at epoch 12\,980 in all nine filters.}
\begin{tabular}{lccc}
\hline\hline
Filter & ER Flux & Ejecta Flux& Center Flux\\
& $(10^{-15}\,\text{erg}\,\text{cm}^{-2}\,\text{s}^{-1}\text{\AA}^{-1})$ & $(10^{-16}\,\text{erg}\,\text{cm}^{-2}\,\text{s}^{-1}\text{\AA}^{-1})$ & $(10^{-17}\,\text{erg}\,\text{cm}^{-2}\,\text{s}^{-1}\text{\AA}^{-1})$ \\
\hline
F275W & $2.614\pm0.003$ & $4.136\pm0.013$ & $2.594\pm0.031$   \\
F280N & $7.194\pm0.015$ & $15.067\pm0.070$ & $8.280\pm0.165$   \\
F336W &  $1.370\pm0.001$ & $1.060\pm0.004$ & $0.749\pm0.010$   \\
F438W &  $1.229\pm0.001$ & $1.075\pm0.004$ & $0.725\pm0.010$   \\
F502N &  $0.768\pm0.001$ & $0.744\pm0.004$ & $0.464\pm0.009$   \\
F555W & $0.772\pm0.001$ & $0.621\pm0.001$ & $0.450\pm0.004$   \\
F625W &  $1.117\pm0.001$ & $0.784\pm0.002$ & $0.573\pm0.004$   \\
F657N &  $6.879\pm0.003$ & $3.030\pm0.007$ & $2.234\pm0.018$   \\
F814W & $0.237\pm0.001$ & $0.290\pm0.001$ & $0.203\pm0.002$   \\
\hline
\end{tabular}
\label{table:flux_2022}
\end{table*} 

\begin{table*}
\centering
\caption{De-reddened flux measurements in the ER, ejecta, and center at epoch 8329 in all eight filters.}
\begin{tabular}{lccc}
\hline\hline
Filter &ER Flux & Ejecta Flux& Center Flux\\
& $(10^{-15}\,\text{erg}\,\text{cm}^{-2}\,\text{s}^{-1}\text{\AA}^{-1})$ & $(10^{-16}\,\text{erg}\,\text{cm}^{-2}\,\text{s}^{-1}\text{\AA}^{-1})$ & $(10^{-17}\,\text{erg}\,\text{cm}^{-2}\,\text{s}^{-1}\text{\AA}^{-1})$ \\
\hline
F225W & $7.402\pm0.015$ & $4.236\pm0.037$ & $3.166\pm0.101$ \\
F336W &  $2.790\pm0.003$ & $0.809\pm0.006$ & $0.641\pm0.017$   \\
F438W & $2.418\pm0.002$ & $0.915\pm0.004$ & $0.692\pm0.012$  \\
F502N & $2.402\pm0.002$ & $0.667\pm0.003$ & $0.479\pm0.009$ \\
F555W & $1.614\pm0.001$ & $0.511\pm0.002$ & $0.411\pm0.006$ \\
F625W & $2.377\pm0.001$ & $0.611\pm0.001$ & $0.509\pm0.002$  \\
F657N & $17.144\pm0.006$ & $2.665\pm0.008$ & $2.286\pm0.024$ \\
F814W & $0.488\pm0.001$ & $0.261\pm0.001$ & $0.200\pm0.003$ \\
\hline
\end{tabular}
\label{table:flux_2009}
\end{table*}

The regions encompassing the ejecta, and, to a lesser extent, the center, are contaminated by scattered ER emission. Given that we cannot separate the ER contribution from the ejecta and center contributions based on the images, we had to rely on a synthetic model of the ER (see Appendix\,\ref{appendix:ER_synthetic}) to remove the contribution of the ER to the ejecta and center. It led to a reduction of the ejecta and center fluxes by 6.3\% to 20.8\%, depending on the date of observation and filter (see Table\,\ref{table:flux_reduction_2022_2009} in Appendix\,\ref{appendix:ER_synthetic}). The ER contribution to the ejecta and center does not change much with time because the fading of the ER and the expanding ejecta and center apertures affect this level in opposite directions.

We additionally note that our line of sight from Earth places the southern rim of the northern outer ring in projection behind the center of the ejecta, adding to its flux. 
This contribution cannot easily be removed due to the spatial variations of the brightness of the outer rings. However, we used the faintest and brightest parts of the northern outer ring outside the ER to place limits on the contribution of the northern outer ring (Appendix\,\ref{appendix:outerrings}). The contributions are $\lesssim 5\%$ in the ejecta region and $\lesssim 15\%$ in the center region for most filters, typically changing by $\lesssim 3\%$ between the epochs. The largest contributions by far are seen in the F502N filter, where it reaches maximal levels of 46\% and 25\% in the center region at epochs 12\,980 and 8329, respectively (see Table\,\ref{table:outerring} in Appendix\,\ref{appendix:outerrings}).

Given that the X-ray input from the ER to the ejecta is not uniform and especially presents an east-west asymmetry \citep{frank16}, it is interesting to investigate the asymmetries between the east and west parts of both the ER and ejecta in the HST data. In order to quantify differences between the east and west regions seen in the images (see Sect.\,\ref{sect:morphology}), we separated the ejecta and ER regions into east and west parts, the boundary separating the two parts being the major-axis of the ellipse encompassing the ejecta region (dashed line in Fig.\,\ref{fig:regions}). We then proceeded as for the full regions to compute the fluxes in the east and west parts. 

The photometry of the full regions at epochs 12\,980 and 8329 is provided in Tables\,\ref{table:flux_2022} and \ref{table:flux_2009}, and presented in Fig.\,\ref{fig:SEDs_2009_and_2022} together with the ratio between the western and eastern parts for the ER and ejecta, as well as the ratio between the center and ejecta. The ratios between the fluxes at epochs 12\,980 and 8329 are presented in Fig.\,\ref{fig:ratio_SEDs_2009_and_2022} for the ER, ejecta, and center regions, as well as the eastern and western parts of the ER and ejecta. Since the shortest wavelength filters are different in the two epochs, they are plotted side by side in Fig.\,\ref{fig:SEDs_2009_and_2022} and not included in Fig.\,\ref{fig:ratio_SEDs_2009_and_2022}. 

The ER is overall dimmer by a factor two at epoch 12\,980 compared to epoch 8329, with the dimming more pronounced in the narrow-band filters. The eastern part has dimmed more than the western part (except in the F336W filter where no difference is observed between east and west).

The ejecta have brightened by $\sim$\,20\% between the two epochs, the largest increases being observed in the F336W and F625W filters. The photometry of the ejecta shows more variations as a function of wavelength than the one of the ER. Overall, the western part of the ejecta has brightened more than the eastern part, except in the F336W filter where no noticeable difference is observed, and except the eastern part in the F657N filter that has slightly dimmed. The center region has brightened in the wide-band filters, less than in the ejecta region, but has not changed significantly in the narrow-band filters. The photometry of the center is very similar to the photometry of the ejecta (with small variations seen in the bottom right panel of Fig.\,\ref{fig:SEDs_2009_and_2022}) and has evolved similarly with time as the photometry of the ejecta. 

The full luminosity of the ER, ejecta, and center in the UV-optical band covered by the HST broad-band filters amount to $(5.75\pm0.01)\times10^{35}$, $(5.69\pm0.02)\times10^{34}$, and $(3.88\pm0.04)\times10^{33}$ in 2022, and $(1.183\pm0.001)\times10^{36}$, $(4.47\pm0.02)\times10^{34}$, and $(3.49\pm0.07)\times10^{33}$\,erg\,s$^{-1}$ in 2009, respectively. These luminosities were computed based on the FWHM of the filters and corresponding fluxes in these filters. Though the filters overlap over certain wavelength ranges, they do so where the ejecta spectrum is free of any dominant line (see Fig.\,\ref{fig:spectra}) so our determination of the total luminosities should not be affected by these overlaps.

\section{Light curves\label{sect:lightcurve}}
This section is devoted to the analysis of the light curves of the ER, ejecta, and center of SN\,1987A in the two filters, F438W and F625W, that are common to all twelve epochs from day 8329 to day 12\,980, and is a major update from \citet{larsson19a}. The fluxes were computed as described in Sect.\,\ref{sect:SEDs}. The observations in the F438W and F625W filters are shown in Fig.\,\ref{fig:center_f438w_2009_to_2022} and Appendix\,\ref{appendix:additionalfigures}, Fig.\,\ref{fig:Ejecta_f625w_2009_to_2022} (optimized to see the ejecta) together with the adopted ejecta and center regions, and Fig.\,\ref{fig:ER_all_2009_to_2022} (optimized to see the ER). Both ejecta and center regions are expanded linearly with time for the assumed homologous expansion of the ejecta following Eq.\,\eqref{eq:ejectavel}.
The ejecta region clearly misses the bright southern part of the ejecta but we cannot make it bigger due to overlap with the ER in recent years. A direct quantitative comparison of the fluxes to previous works is impossible as we adopted a smaller region size for the ejecta to encompass most of the ejecta in the latest observations while avoiding the ER. 

Depending on the orientation of the telescope at the time of the observation, diffraction spikes from Stars 2 and 3 might cross the center, ejecta, or ER regions\footnote{The observations at epochs 8329 and 12\,980 were not affected by diffraction spikes from Stars 2 and 3.}. The extra contribution of the diffraction spikes to the flux in the corresponding region was subtracted as described in Appendix\,\ref{appendix:diffraction_spikes}. This correction is at the level of 4\% at most. We also modeled and removed the contribution of the scattered light from the ER to the ejecta and the center, as explained in Appendix\,\ref{appendix:ER_synthetic}. The contribution of the northern outer ring to the ejecta and center (see Fig.\,\ref{fig:general}) amounts to $0-7$\% and $1-17$\% in the F438W filter and $1-4$\% and $3-10$\% in the F625W filter (estimated as described in Sect.\,\ref{sect:SEDs} and Appendix\,\ref{appendix:outerrings}), respectively.

\begin{figure}
\centering
\includegraphics[clip=true,trim=0 0 0 0,width=1\linewidth]{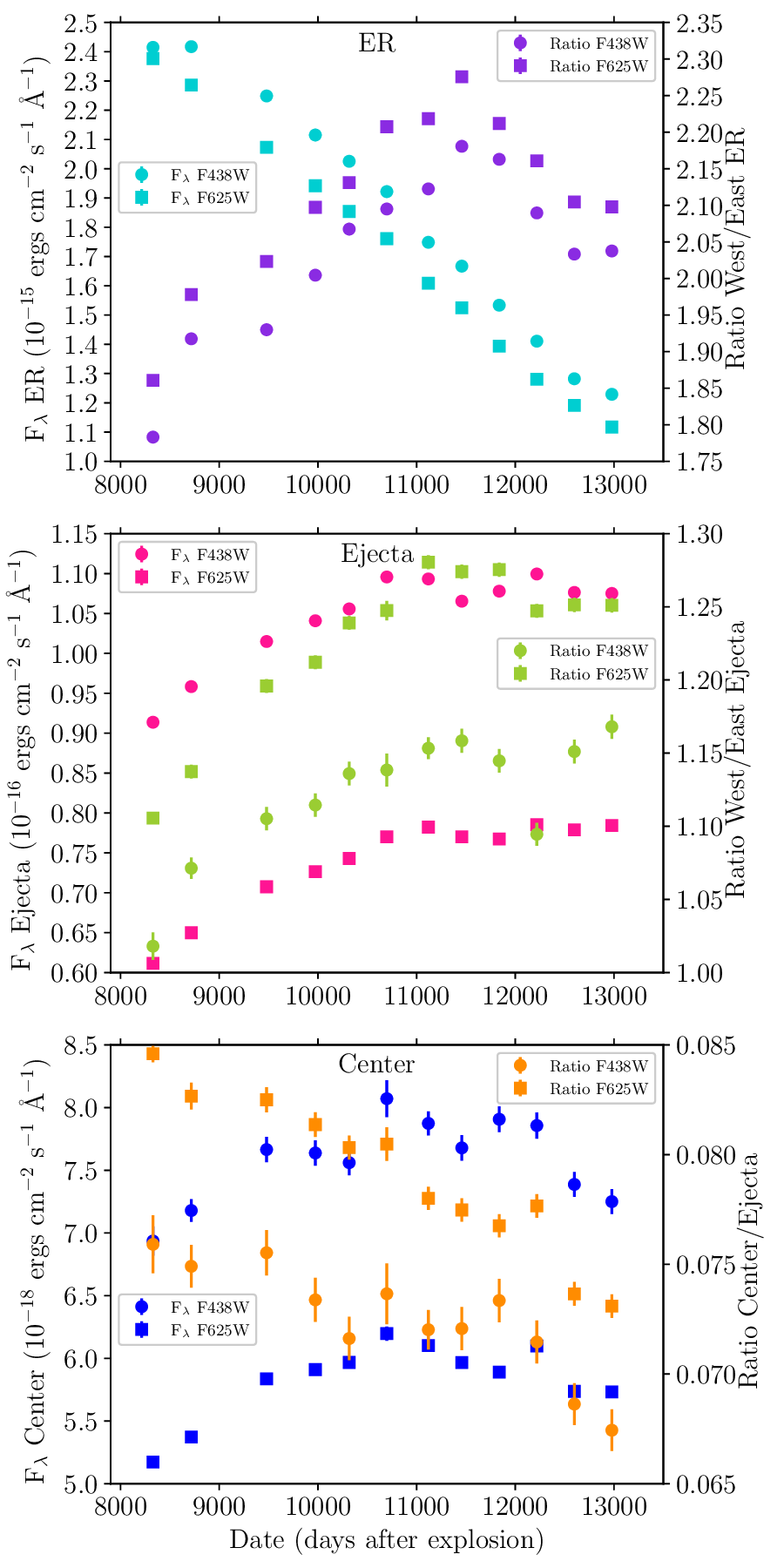}
\caption{\textit{Top panel:} Light curve of the ER and ratio between west and east parts. \textit{Middle panel:} Same for the ejecta. \textit{Bottom panel:} Light curve of the center and ratio between center and ejecta fluxes. We note that the error bars are smaller than the symbols in most cases.\label{fig:lightcurve_2009_to_2022}}
\end{figure}

The flux measurements in the ER, the ejecta, and the center between epochs 8717 and 12\,598 in the F438W and F625W filters are provided in Table\,\ref{table:flux_2011_to_2021}. The light curves are presented in the three panels of Fig.\,\ref{fig:lightcurve_2009_to_2022} for the ER, ejecta, and center together with the ratio between the western and eastern parts for the ER and ejecta and the ratio between the center and the ejecta. Light curves for the ER and ejecta extending to earlier epochs are presented in \citet[their figures 10 and 11]{larsson19a}. The exact flux values are not directly comparable with this work due to the above-mentioned small differences in apertures, and the fact that correction factors were applied to the light curves in \citet{larsson19a} to account for the differences between the three different instruments used (WFPC2, ACS, and WFC3). However, the general time evolution is not affected by these differences, and follows a similar trend in both works, therefore enforcing that adopting different apertures does not change the results qualitatively.

\begin{table*}
\centering
\caption{De-reddened flux measurements in the ER, ejecta, and center between epochs 8717 and 12\,598 in the F438W and F625W filters.}
\begin{tabular}{lccc}
\hline\hline
Epoch (days) & ER Flux & Ejecta Flux& Center Flux\\
& $(10^{-15}\,\text{erg}\,\text{cm}^{-2}\,\text{s}^{-1}\text{\AA}^{-1})$ & $(10^{-16}\,\text{erg}\,\text{cm}^{-2}\,\text{s}^{-1}\text{\AA}^{-1})$ & $(10^{-18}\,\text{erg}\,\text{cm}^{-2}\,\text{s}^{-1}\text{\AA}^{-1})$ \\
\hline
\multicolumn{4}{c}{Filter F438W}\\
8717 & $2.417\pm0.002$ & $0.958\pm0.003$ & $7.179\pm0.090$\\
9480 &  $2.249\pm0.002$ & $1.015\pm0.004$ & $7.665\pm0.101$\\
9974 & $2.115\pm0.002$ & $1.041\pm0.004$ & $7.638\pm0.101$\\
10\,317 &  $2.026\pm0.002$ & $1.056\pm0.004$ & $7.561\pm0.101$\\
10\,698 & $1.922\pm0.002$ & $1.096\pm0.005$ & $8.071\pm0.147$\\
11\,119 & $1.749\pm0.001$ & $1.093\pm0.004$ & $7.874\pm0.095$\\
11\,458 & $1.667\pm0.001$ & $1.065\pm0.004$ & $7.679\pm0.101$\\
11\,837 & $1.534\pm0.001$ & $1.078\pm0.004$ & $7.907\pm0.103$\\
12\,218 & $1.411\pm0.001$ & $1.099\pm0.004$ & $7.857\pm0.104$\\
12\,598 & $1.282\pm0.001$ & $1.076\pm0.004$ & $7.387\pm0.099$\\
\multicolumn{4}{c}{Filter F625W}\\
8717 &  $2.286\pm0.001$ & $0.650\pm0.001$ & $5.372\pm0.038$\\
9480 &  $2.073\pm0.001$ & $0.707\pm0.001$ & $5.836\pm0.039$\\
9974 & $1.942\pm0.001$ & $0.726\pm0.001$ & $5.910\pm0.039$\\
10\,317 &  $1.854\pm0.001$ & $0.743\pm0.001$ & $5.968\pm0.039$\\
10\,698 & $1.761\pm0.001$ & $0.770\pm0.002$ & $6.198\pm0.056$\\
11\,119 & $1.609\pm0.001$ & $0.782\pm0.001$ & $6.102\pm0.040$\\
11\,458 & $1.525\pm0.001$ & $0.770\pm0.001$ & $5.967\pm0.039$\\
11\,837 & $1.393\pm0.001$ & $0.767\pm0.001$ & $5.890\pm0.039$\\
12\,218 & $1.280\pm0.001$ & $0.785\pm0.001$ & $6.098\pm0.040$\\
12\,598 & $1.191\pm0.001$ & $0.779\pm0.002$ & $5.737\pm0.041$\\
\hline
\end{tabular}
\label{table:flux_2011_to_2021}
\end{table*} 

The decline in the light curve of the ER is explained by the fading of the hotspots that started $\sim 8000$\,days after the explosion \citep{fransson15}. The fading of the hotspots is interpreted as the forward shock (with light-travel delays) reaching the full circumference of the ER, such that no new hospots were appearing, and the older hotspots fading as they were destroyed by the shock. In both filters, the ratio between the fluxes in the western and eastern parts of the ER increased until day $\sim11\,500$, after which it decreased. This indicates that the eastern part has proportionally faded more and more compared to the western part until day $\sim11\,500$, after which the fading rate of the western part started to catch up with that of the eastern part. This could be due to the increasing contribution of the reverse shock emission in the east, which falls inside the defined aperture for the ER. 

An increase in the ejecta flux is observed until day $\sim10\,700$ (F438W) and $\sim11\,100$ (F625W), after which the flux flattens at a constant value. The ratio between the fluxes in the western and eastern parts of the ejecta in the F625W and F438W filters follows the increase observed in the total flux in the corresponding filter until day $\sim\,11\,100$, after which it drops to values of 1.25 and 1.16, respectively. This shows that most of the brightening between epochs 8329 and 12\,980 occurred before $\sim10\,700$\,days. 
The small, though noticeable, increase in the ejecta flux in the two filters at day $12\,200$, as well as the drop at the same date in the ratio between the fluxes in the western and eastern parts of the ejecta in the F438W filter, are unlikely to have any astrophysical origin but are rather due to an improper diffraction spike contribution estimate. 
Indeed, this level of variability on a one year time scale is unlikely to be real. 
The color of the ejecta, determined from the flux ratio between the F625W and F438W filters, has stayed roughly constant (values between 0.66 and 0.72 in unitless ratios) between days 8329 and 12\,980, therefore supporting that the energy source and physical conditions have not changed in this time interval.

The light curve of the center region shows a clear increase across the first three epochs, followed by a general flattening after day $\sim10\,000$ (see Fig.\,\ref{fig:lightcurve_2009_to_2022}). The ratio between center and ejecta fluxes has overall decreased with time, highlighting the increasing flux in the outer part of the ejecta (as best seen in Fig.\,\ref{fig:2022_shrinked_minus_2009}), pointing toward an external source of energy powering the outer parts of the inner ejecta in addition to the $^{44}\text{Ti}$ radioactivity, notably X-rays from the ER. 

\section{Discussion\label{sect:discussion}}
SN\,1987A is the only modern SN for which the time evolution of the ejecta can be studied with spatial resolution -- providing information about the evolving physical conditions and revealing the explosion geometry in increasingly great detail. 
For the range of epochs considered in this paper (from day 8329 to day 12\,980 after the explosion), the ejecta have expanded by a factor of 1.56, allowing us to resolve correspondingly smaller spatial scales. The observed ejecta morphology is strongly affected by the various time-evolving energy sources, so we start by discussing them in  Sect.\,\ref{subsect:energysources}, followed by a discussion of the line excitation mechanisms and their connection to the ejecta morphology in Sect.\,\ref{subsect:lineexcitation}. We then discuss the intrinsic, asymmetric distribution of the ejecta in Sect.\,\ref{subsect:asymmetricdistribution}, and finally consider the implications of the non-detection of the compact object in the HST images in Sect.\,\ref{subsect:compactobject}.
 
\subsection{Energy sources in the ejecta\label{subsect:energysources}}
Previous analysis of the ejecta light curve from HST imaging showed that the ejecta faded as expected from the decay of $^{44}$Ti until $\sim$\,$5000$\,days, after which it started brightening. Modeling showed that the brightening is consistent with energy input from X-rays originating from the ER \citep{larsson11,fransson13}.  The flux more than doubled in all filters covering the $2000-11000$\,\AA~ wavelength range by day 8329 \citep{larsson11, larsson13}, clearly making X-rays the dominant energy source at late times. The fluxes measured in the F625W and F438W filters subsequently continued to increase, but at a gradually lower rate, showing signs of levelling off at day $\sim$\,$11\,000$ \citep{larsson19a}. Here we have added four more epochs of data compared to \cite{larsson19a}, which confirm the flattening of the light curves and show that the ejecta flux has remained approximately constant since $\sim$\,$11\,000$\,days (see Fig.\,\ref{fig:lightcurve_2009_to_2022}, middle). Considering the full wavelength range, the brightening between days 8329 and 12\,980 is $\sim$\,$20$\% (see Fig.\,\ref{fig:ratio_SEDs_2009_and_2022}).

\begin{figure}
\centering
\includegraphics[width=1\linewidth]{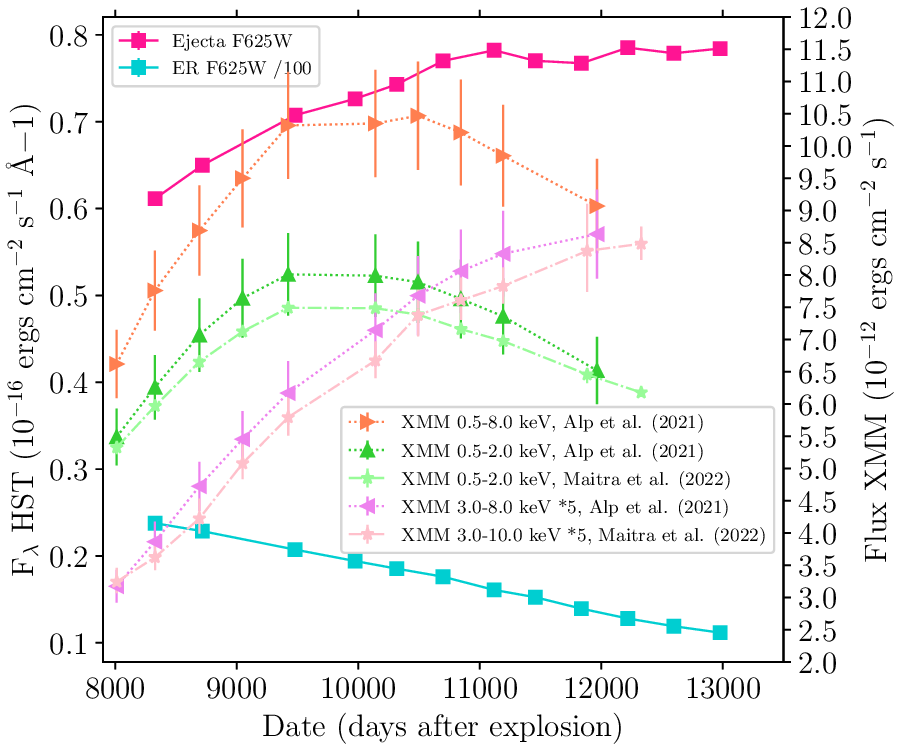}
\caption{X-ray light curves of the ER from XMM-Newton in different energy bands taken from \citet{alp21} and \citet{maitra22} together with the light curves of the ER and the ejecta in the F625W filter. Some light curves were scaled by a factor of 5 for clarity purposes (see legend).\label{fig:XMM}}
\end{figure}

Figure~\ref{fig:XMM} shows the HST light curves of the ejecta and ER in the F625W filter together with light curves of the X-ray emission from the ER in different energy bands from XMM-Newton observations \citep{alp18, maitra22}. The soft X-ray ($0.5-2$\,keV) light curve started decreasing after day $\sim$\,$10\,000$, while the hard X-ray ($3-8$\,keV) light curve shows signs of flattening around day $\sim$\,$12\,000$. These transitions in the X-ray light curves occur around the same epoch as the optical light curves of the ejecta start flattening.

However, a one-to-one relation between X-rays from the ER and optical ejecta emission is not expected, considering that the flux evolution of the ejecta is affected by the light-travel time between the relative positions of the expanding ejecta and the X-ray emission in 3D, by the decreasing ejecta density, as well as by the evolving shape of the X-ray spectrum \citep{fransson13}. 
 
The energy input from the X-rays also affects the observed UV-optical morphology of the ejecta. This is most clearly seen from the fact that the western part of the ejecta has brightened more than the eastern part between days 8329 and 12\,980 (see Fig.\,\ref{fig:2022_shrinked_minus_2009}), the total optical flux in the western part being a factor 1.2 brighter in the latest epoch (Fig.\,\ref{fig:SEDs_2009_and_2022}, lower middle panel). This is compatible with the fact that the western part of the ER is brighter than the eastern part by $\sim$\,$1.6$ in the X-rays at 10\,433 days after the explosion, resulting in more energy being deposited in the western ejecta \citep[see figure~6 in][]{frank16}.

In addition to affecting the west/east flux ratios, the X-ray input  leads to a limb-brightened morphology of the ejecta. Most of the X-rays are expected to be absorbed in the outer part of the inner ejecta (the edge-brightened region seen in Fig.\,\ref{fig:ejecta_all_2022}), given the increase in metallicity and the steepening of the density gradient at the boundary of the ejecta core (see models in \citealt{fransson13}). This is supported by our observations, which show that the central region of the ejecta has brightened less than the ejecta as a whole, and even decreased in brightness in the narrow-band filters (see Fig.\,\ref{fig:ratio_SEDs_2009_and_2022}). 
In summary, the light curves, east/west asymmetry, and limb-brightened morphology all support the general scenario that X-rays from the ER dominate the energy input to the ejecta. By contrast, energy input from $^{44}$Ti is instead expected to result in a gradual fading without significant spatial variations.

The only part of the ejecta where $^{44}$Ti may still dominate is the innermost region.
Indeed, the “hole” in the ejecta located just north of the center \citep[e.g.,][]{larsson11, larsson13} is still clearly visible (see e.g., Fig.\,\ref{fig:ejecta_all_2022}), and Fig.\,\ref{fig:2022_shrinked_minus_2009} indicates that it has faded between epochs 8329 and 12\,980 in almost all filters. 
To quantify the possible fading of the “hole” we integrated the flux over a region as large as the region adopted for the center, but centered at the middle of the hole rather than at the center of the ER. We found that the flux  decreased by $\sim$\,7\% in the filters which show a clear dimming of the hole (see Sect.\,\ref{subsect:morphologies_evolution}).
Contamination in the “hole” region by the bright surrounding ejecta is likely non-negligible, especially in the smaller region at day 8329, and might therefore impact our determination of the flux. Nevertheless, it is interesting to compare these results with the fading expected between these epochs for a region of ejecta powered only by $^{44}$Ti radioactivity, 
\begin{equation}
\Delta = \left(1 - \exp\left(\frac{t_2-t_1}{\tau}\right)\right)\left(\frac{t_1}{t_2}\right)^2,
\end{equation}
which is 7\% for the bolometric flux between $t_1 = 8329$ days and $t_2=12\,980$ days, assuming a linear expansion and a mean lifetime $\tau$ of 85 years. Without the IR fluxes the bolometric flux cannot be computed, but the evolution of the HST fluxes is compatible with the interpretation that $^{44}$Ti dominates over the X-ray input in the “hole”. 

Finally, the interpretation of the flux evolution in the central ejecta is further complicated by a large amount of dust, the detailed properties of which are uncertain \citep[see e.g.,][]{matsuura22}. An important consideration regarding the dust is that the optical depth is expected to decrease as the inverse of the time squared. The fact that we do not observe any strong brightening in the central region points toward dust still highly optically thick. 

\subsection{Line excitation and connection to the morphology\label{subsect:lineexcitation}}
The X-rays coming from the ER are thermalized, which means that their photoabsorption generates free electrons that ionize and excite atoms in the ejecta. The hydrogen and iron-like emission lines are then produced by recombination. This process results in a similar flux increase in the ejecta and west/east ratio in many of the broad-band filters (see Figs\,\ref{fig:ejecta_all_2022} and \ref{fig:SEDs_2009_and_2022}). Two notable differences stand out from this picture: 1) The [Ca\,\textsc{ii}] lines contributing to the F814W filter show a smaller increase in flux, and 2) the Mg\,\textsc{ii} lines contributing to the UV filters show a high increase in flux and also a very different morphology compared to the other filters.

The smaller increase in flux in the  [Ca\,\textsc{ii}]-dominated F814W filter (10\%) compared to the other broad-band filters ($20-30$\%) is significant (see Figs\,\ref{fig:ratio_SEDs_2009_and_2022} and \ref{fig:SEDs_2009_and_2022}, upper middle panel). A likely explanation for this is that a significant fraction of the [Ca\,\textsc{ii}] emission comes from excitation through the resonant H \& K lines at 3934 and 3968\,\AA\ by fluorescence, which one decreases over time \citep{li93, kozma98}. In contrast to the earlier epochs, where the flux in the H \& K lines was dominated by the ejecta, the flux is now dominated by radiative shocks in the ER \citep[see][and references therein]{fransson13}. Scattering by the H \& K lines gives rise to the triplet of Ca\,\textsc{ii} lines at 8498, 8542, and 8662\,\AA, which in turn gives rise to the  [Ca\,\textsc{ii}] $\lambda\lambda$\,7293, 7326 emission lines. Therefore, the emissions in the doublet and triplet should be of the same order. The triplet is not observed in the STIS spectra (see Fig.\,\ref{fig:spectra}) because of the low S/N of the spectra in this wavelength range. However, it is observed in newly acquired MUSE spectra that will be discussed in a forthcoming paper (Fransson et al., in prep.), which supports this interpretation.  The fact that the contribution from this process has decreased with time -- as the ER has faded by a factor of $\sim 2$ between the two considered epochs -- explains that the flux in the F814W filter has increased less than the flux in the hydrogen-dominated filters.

For the Mg\,\textsc{ii} emission that dominates the F275W and F280N filters, the ejecta morphology is rather different than the morphology in the other filters. Notably, the Mg\,\textsc{ii} emission is brighter in the western and eastern regions between the inner ejecta and ER, fainter in the southern ejecta close to the ER, and fainter in the central region just south of the hole (see Figs\,\ref{fig:ejecta_all_2022} and \ref{fig:ratio_images}). A major reason for this is that the Mg\,\textsc{ii} $\lambda\lambda$\,2795, 2802 lines are optically thick, so their emission indicates where they are last scattered rather than emitted. \cite{kangas22} found that these lines are primarily powered by X-rays from the ER (which can explain the bright emission in the western ejecta, as discussed for all the filters in Sect.\,\ref{subsect:energysources}) but with a significant ($\sim 1/3$) contribution from  pumping of the Mg\,\textsc{ii} 1239, 1240\,\AA\ transition by redshifted Ly\,$\alpha$ (1216\,\AA).

Given the difference in wavelength with the Mg\,\textsc{ii} $\lambda$\,1240 line, the Ly\,$\alpha$ line responsible for the pumping cannot be emitted from the same location as the Mg\,\textsc{ii} line, but has to come from a region receding from the Mg\,\textsc{ii} location with a relative velocity between 5981 and 6097\,km\,s$^{-1}$, corresponding to a redshift of $24.26-24.73$\,\AA. Because of the damping of the wings of the Ly\,$\alpha$ line \citep[see figure~4 and related discussion in][]{kangas22}, these values are upper limits only as a non-negligible part could come from longer wavelength Ly$\alpha$ photons, that is to say, lower redshift photons.

Given the expansion of the ejecta, the Ly\,$\alpha$ photons important for the Mg\,\textsc{ii} emission are emitted by close to diametrically opposed regions\footnote{In comparing the Ly\,$\alpha$ emission with the Mg\,\textsc{ii} emission, we should account for the temporal delay between the two emissions. For the ejecta region considered here, the maximum delay corresponds to the diameter of the ER divided by the speed of light, which is equal to 1.7 years. However, the effect of this time delay is negligible for the comparisons that can be made based on the imaging data.}. The absence of Mg\,\textsc{ii} emission in the central part of the ejecta most probably implies that there is no such Ly\,$\alpha$ emission region having the correct redshift to produce the Mg\,\textsc{ii} emission. Another explanation (not exclusive) could be that the region is optically thick to the Ly\,$\alpha$ photons, which prevents the Mg\,\textsc{ii} emission in this central region. This was indeed found by \citet{jerkstrand11} at 2875\,days after the explosion but since then the optical depth has gone down over time because of the expansion. 

The obvious candidate for the origin of the Mg\,\textsc{ii} line emission in the inner ejecta is the Ly\,$\alpha$ emission in the reverse shock, as the latter is known to emit strong Ly\,$\alpha$ and to extend to velocities of $\sim10\,000$\,km\,s$^{-1}$ \citep{france11}. Further, the reverse shock emission is mainly blueshifted in the north and redshifted in the south \citep[see e.g.,][]{france15, kangas22, larsson23}.

Figure~5 of \citet{larsson23} shows that the regions having the highest radial velocities 12\,927 days after the explosion are located in the southeast reverse shock, which is consistent with the highest Mg\,\textsc{ii} emission arising in the western part of the ejecta. This is also illustrated in figures 15 and 16 of \citet{larsson19a} showing contours in velocities in the reverse shock in agreement with the previous assertion.

According to \citet{michael03}, the Ly\,$\alpha$ photons from the reverse shock will not easily traverse the ER back and propagate to the center. However, the photons that undergo scattering towards the center will be redshifted by the correct amount \citep[up to a velocity of 12\,000\,km\,s$^{-1}$,][]{michael03, larsson19a} and contribute to the Mg\,\textsc{ii} emission, consistent with \citet{kangas22}.

In addition, the N\,\textsc{v} $\lambda\lambda$\,1239, 1243 lines emitted by the reverse shocks might also contribute to the Mg\,\textsc{ii} fluorescence, though to a lower extent than the Ly\,$\alpha$ emission, as discussed in \citet{kangas22}. Given the wavelengths of the N\,\textsc{v} lines, these lines should be emitted in a region of the reverse shocks that have no (or very small) velocity shift compared to the considered part of the ejecta. 
We therefore conclude that the conditions are met for the Ly\,$\alpha$ and N\,\textsc{v} lines to contribute to the observed Mg\,\textsc{ii} emission. 

\subsection{Asymmetric distribution of the ejecta\label{subsect:asymmetricdistribution}}
The asymmetric distribution of the ejecta observed at late times carries the imprints of the explosion mechanism and progenitor structure, as shown by numerical simulations \citep{orlando20, ono20, gabler21}. The best observational constraints on the ejecta morphology in SN\,1987A have been obtained from integral field unit data and long-slit spectroscopy, which provide 3D information for individual emission lines \citep{kjaer10, larsson16, larsson19b, larsson23, kangas22}. In contrast, the morphology observed in images is more difficult to interpret due to projection effects, blending of different emission lines in broad-band filters, as well as incomplete coverage of the broad ejecta lines in the narrow filters. The HST images analyzed here do, however, provide an important complement to the 3D results by providing detailed information about the time evolution, better spatial resolution, as well as a wider wavelength coverage.  

Figures~\ref{fig:center_f438w_2009_to_2022} and \ref{fig:Ejecta_f625w_2009_to_2022} illustrate the time evolution of the ejecta and includes the expanding aperture used for measuring the ejecta flux. The outer edges of the bright inner ejecta remain at the same relative position with respect to this aperture, as expected for homologous expansion (e.g., the bright extension to the west that just reaches the edge of the aperture at all epochs). 

\begin{figure*}[h]
\centering
\includegraphics[clip=true,trim=20 20 0 0,width=1\linewidth]{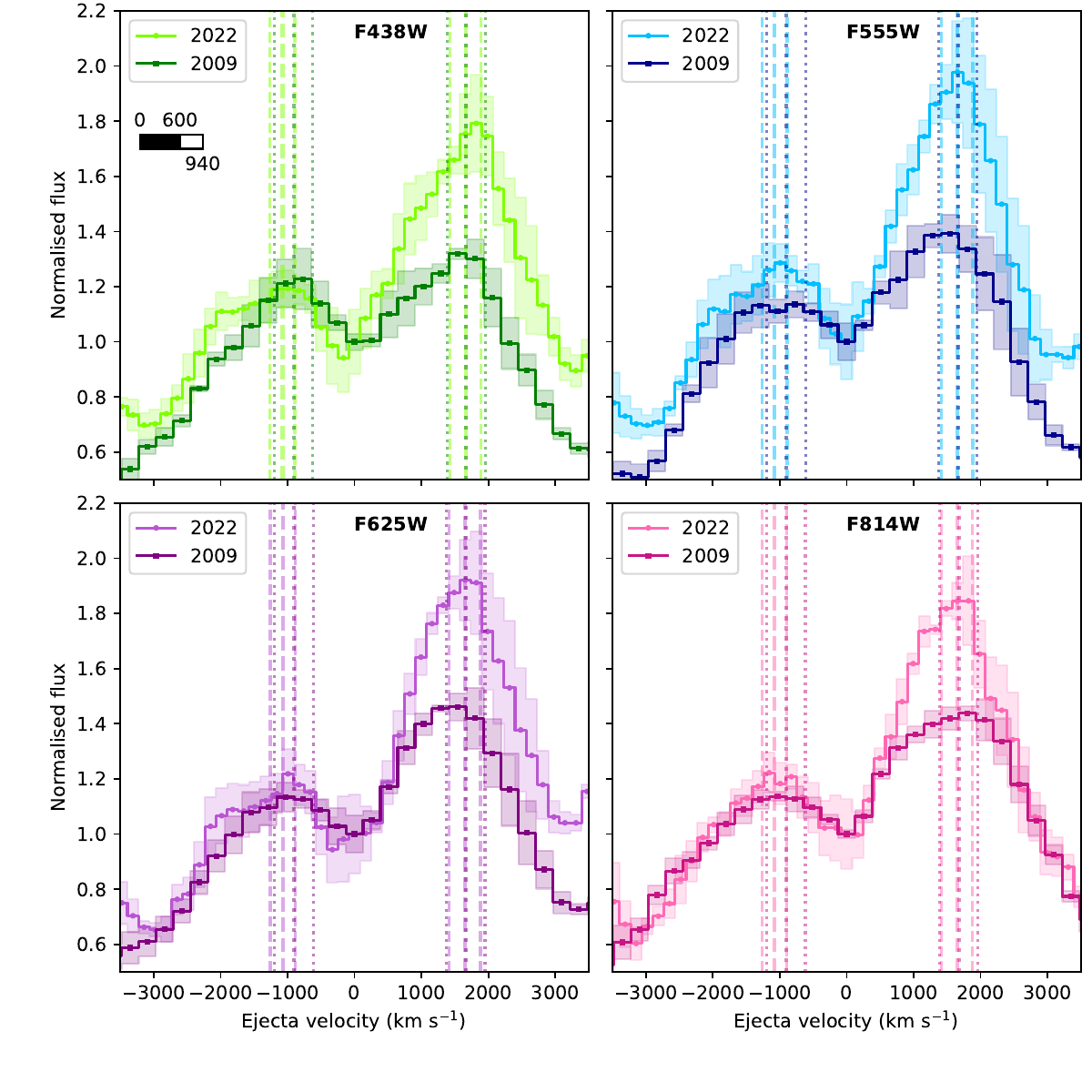}
\caption{Normalized flux as a function of the position, converted into ejecta velocity with respect to the geometric center,  in the east-west region, in 2022 and 2009 in the F438W, F555W, F625W, and F814W filters. The lines and shaded areas correspond to the median fluxes and corresponding standard deviations. The vertical dashed (resp. dotted) lines indicate the mean and standard deviation of the maxima of the corresponding 2022 (resp. 2009) curves. The scale bars in the top left panel indicate the velocity resolution corresponding to the FWHM of the spatial resolution at the two epochs.\label{fig:homologous_expansion}}
\end{figure*}

To quantify the homologous expansion, we adopted a horizontal region centered on the geometric center of SN\,1987A covering the inner ejecta from east to west and with a width in the north-south direction of $0\fds15$ in the 2022 observations (see Fig.\,\ref{fig:regions}). For each position in the east-west direction, we computed the median flux and associated standard deviation over the pixels in the north-south direction. We adopted the same procedure for the 2009 observations except that the width in the north-south direction was scaled by the ratio between days since explosion to account for the homologous expansion. The results are shown for four different filters in Fig.\,\ref{fig:homologous_expansion}, where the position in the east-west direction has been converted into expansion velocity following Eq.\,\eqref{eq:ejectavel}. In these plots, the fluxes are in arbitrary units and were scaled by the corresponding flux in the central pixel for clarity. As  Fig.\,\ref{fig:homologous_expansion} shows, the two peaks, corresponding to the two brighter regions east and west of the center, are located at positions corresponding to expansion velocities of $-1075\pm185$ and $+1650\pm230$\,km\,s$^{-1}$ in 2022, and $-905\pm285$ and $+1670\pm285$\,km\,s$^{-1}$ in 2009, corresponding to the mean and standard deviation in all four filters presented in Fig.\,\ref{fig:homologous_expansion}. The fact that the peaks occur at very similar velocities at both epochs, consistent within $1\sigma$ strongly supports that the ejecta is in homologous expansion. The results for the other filters as well as for a region oriented in the north-south direction are very similar.

The main time evolution observed is the relative brightening of the western side compared to the eastern side, attributed to energy input from X-rays from the ER (see Sect.\,\ref{subsect:energysources}). We note that the expansion implies that the spatial resolution (FWHM $\sim$ 0\farcs{09}), expressed in terms of ejecta velocity following Eq.\,\eqref{eq:ejectavel}, has improved from $\sim940$ to $\sim600$\,km\,s$^{-1}$ over the epochs covered by Fig.\,\ref{fig:center_f438w_2009_to_2022}. While we observe spatial variations in the surface brightness down to the resolution level of WFC3 at all epochs (also illustrated in Fig.\,\ref{fig:homologous_expansion}), no new clearly separated substructures have emerged over the course of the observations. This indicates that there are no well-defined clumps on this size scale or that any such clumps are hidden by projection effects in the images.

The comparison of the ejecta morphology at different wavelengths (see Fig.\,\ref{fig:ejecta_all_2022}) reveals the same overall elongation along an axis offset to the east from the north by $\sim 15^{\circ}$. This shows that the elements that dominate the thermal emission in different filters, that is to say, Mg (F275W), Fe (F336W and F438W), H (F555W and F625W), and Ca (F814W), are well mixed and follow the same large-scale geometry as has been seen in previous observations \citep[and references therein]{larsson23}. The differences between filters that are observed on a more detailed level are expected to reflect both smaller-scale spatial differences in ejecta abundances and the details of the emission line production. While the morphology of the Mg emission is clearly influenced by scattering effects (see Sect.\,\ref{subsect:lineexcitation}), it is interesting to compare the filters dominated by Fe, H, and Ca. The corresponding images all show enhanced emission in the western ejecta. However, the H-dominated filters are relatively brighter in the region that extends furthest to the west near the mid-plane of the ER (see Figs\,\ref{fig:ejecta_all_2022} and \ref{fig:ratio_images}), which is consistent with a radially increasing relative abundance of hydrogen.

\subsection{Constraints on the compact object}\label{subsect:compactobject}
JWST observations of SN 1987A in the NIR and mid-IR showed strong evidence for a central source in emission lines of [S\,\textsc{iii-iv}], [Ar\,\textsc{ii}], and [Ar\,\textsc{vi}] \citep{fransson24}. The lines are narrow ($\lesssim300$\,km\,s$^{-1}$) and the source is spatially unresolved, showing that the emission region is small. The small, central emission region, the high ionization, and the fact that these lines were only seen in these intermediate elements indicate that this emission is coming from the explosive oxygen burning region, close to the newly formed neutron star. The exact nature of this emission is, however, not clear. The strongest candidates for the ionizing emission are either synchrotron emission from a pulsar wind nebula (PWN), formed as a result of the relativistic wind from the pulsar, or as a result of the ionizing radiation from the cooling, hot neutron star (CNS), or possibly from slow shocks as a result of the PWN bubble expanding into the ejecta. Photoionization models, reproducing the JWST lines, showed agreement with either the PWN model or the CNS model \citep{fransson24}. 

While these models were only discussed in the context of the JWST observations, they also predict other lines in the UV, optical, and far-IR. In particular they can be used to estimate the fluxes from lines in the observed HST filters. The strongest lines in these models are the [S\,\textsc{ii}] $\lambda\lambda$\,4070, 4076, 6718, 6733, [S\,\textsc{iii}] $\lambda\lambda$\,9071, 9533, [O\,\textsc{iii}]  $\lambda\lambda$\,4960, 5008, [Ar\,\textsc{iii}] $\lambda\lambda$\,7138, 7753, and  [Ar\,\textsc{v}] $\lambda\lambda$\,6437, 7007 lines. The main problem  is the background from the general ejecta emission, as well as that of the outer rings and the reverse shock. Because of the expected line fluxes and especially the filter widths the best prospect for detecting a compact object is the narrow F502N filter, containing the [O\,\textsc{iii}] $\lambda$\,5008 line. 

Using these models and tying them to the JWST observed fluxes, we can estimate the [O\,\textsc{iii}] $\lambda$\,5008 flux. The total luminosity of the strong [Ar\,\textsc{ii}] $6.998\,\mu$m line is $\sim$\,$2.47\times10^{32}$\,erg\,s$^{-1}$, corresponding to a flux of $\sim$\,$8.39\times10^{-16}$\,erg\,cm$^{-2}$\,s$^{-1}$. In the “standard” PWN model \citep[figure 4 in][]{fransson24} the luminosity of the [O\,\textsc{iii}] $\lambda$\,5008 line is $4.8\times10^{-1}$ of the [Ar\,\textsc{ii}] line, while that in the CNS model is much lower, $1.7\times10^{-2}$ of the [Ar\,\textsc{ii}] line. The fluxes therefore correspond  to $\sim$\,$4.03\times10^{-16}$\,erg\,cm$^{-2}$\,s$^{-1}$ and $\sim$\,$1.43\times10^{-17}$\,erg\,cm$^{-2}$\,s$^{-1}$, respectively, in the two models.

There is no point source detected in the F502N filter, so we placed an upper limit on the detection of a point source by adding artificial point sources using the \texttt{DAOStarFinder} routine from the \texttt{photutils.detection} package in \texttt{python}.  We assumed a Gaussian PSF with a FWHM determined from nearby stars. We used a Monte Carlo technique to generate point sources within 3$\sigma$ of the position of the [Ar\,\textsc{vi}] source, which has the most accurate position of the lines detected by JWST, being located $38\pm22$\,mas east and $31\pm22$\,mas south of the center of the ER (with 1$\sigma$ uncertainties, \citealt{fransson24}). Our results show that the highest 3$\sigma$ limit within this region is $6.5\times10^{-18}$\,erg\,cm$^{-2}$\,s$^{-1}$, which we take as the limit on a [O\,\textsc{iii}] emitting point source associated with the JWST source. Considering all the limits in the region, we find that sources of $3.6\times10^{-18}$ to $6.5\times10^{-18}$\,erg\,cm$^{-2}$\,s$^{-1}$ have at least 90\% probability of detection, and sources fainter than $0.9\times10^{-18}$\,erg\,cm$^{-2}$\,s$^{-1}$ have a zero probability to be detected. The upper limits on a point source being lower than both PWN and CNS model predictions.

There are several caveats with this comparison. First, the model fluxes are sensitive to the ionizing spectrum, as is clearly seen when comparing these two models. This is positive when using these lines as diagnostics of the ionizing spectrum and therefore the general scenario. However, there are uncertainties in  the PWN spectrum, where the synchrotron power law of a very young PWN may differ from that of the Crab PWN assumed here, as well as in the temperature of the neutron star. 

A major uncertainty is the abundances of elements in the zone with high S and Ar abundances, which can be seen in figure S7 in \citet{fransson24}. For the 19 M$_\odot$ ZAMS model this extends from 1.77 to 2.10 M$_\odot$. However, while the S and Ar abundances are high over this whole zone, the O abundance is $\lesssim 10^{-4}$ (in number) in the inner 0.11 M$_\odot$ of this zone, while it increases to $\sim 0.6$ in the outer 0.21 M$_\odot$ of the zone. Because the JWST observations are mainly sensitive to the heavier elements, it is difficult to distinguish these zones solely from these observations. However, the different O abundances in the two regions produce widely different results for the [O\,\textsc{iii}] line. The non-detection of this line can therefore be an indication that the emission seen with JWST originates from the inner of these S and Ar rich zones.

Alternatively, the dust in the ejecta might affect the observed flux. The distribution, composition, and optical depth of this dust are uncertain \citep[see][]{fransson24}. For pure silicates the absorption in the optical and NIR ranges is low, but scattering might be important, depending on the size of the grains \citep[e.g.,][]{tamanai17}. The effect of scattering is to both give a spatially more extended distribution and broaden the line in wavelength, if there are multiple scatterings. The total flux would, however, not be affected. However, if the silicates contain even low fractions of other elements, like carbon or iron, the absorption part of the refractive index increases and absorption might become important also for the optical range \citep{dorschner95}. 

In the scenario where pure scattering spreads out the photons of the point source but does not affect the total flux, it is interesting to compare the model predictions with the total flux in the central region in the F502N filter. The measured flux density is  $0.46\times10^{-17}$\,erg\,cm$^{-2}$\,s$^{-1}$\,\AA$^{-1}$ (see Table\,\ref{table:flux_2022}), which corresponds to a flux of $\sim 3.0 \times  10^{-16}$\,erg\,cm$^{-2}$\,s$^{-1}$ for an effective width of 65\,\AA.  The predicted flux from the PWN model in the [O\,\textsc{iii}] line is therefore 134\% of the total, while it is only 4.76\% for the CNS model. If we subtract the maximum possible contamination from the northern outer ring in this region (see Appendix\,\ref{appendix:outerrings}), the predicted fluxes correspond to 249\% and 8.81\%, respectively. This indicates that only the CNS model is consistent with the observations if the assumption of a point source is removed, with the caveat that we do not know the size of the region over which the flux will be distributed due to scattering. 

The contribution from the lines associated with the compact object to the other filters is expected to be smaller than in F502N, so it is not surprising that the photometry of the central region does not reveal any evidence of an additional energy source, showing the same shape but a smaller flux increase with time compared to the surrounding ejecta (see Fig.\,\ref{fig:ratio_SEDs_2009_and_2022}). Similarly, as the light curves in the F438W and F625W filters contain strong, broad lines from the ejecta, it is not surprising that they do not show any increase in the central region due to the compact object. 

\section{Conclusion\label{sect:conclusion}}
SN\,1987A is the most thoroughly studied SN so far thanks to the spatially resolved view of the transition into a young SNR that it offers.
In this paper, we have analyzed the HST images taken at day 12\,980 after the explosion. These new images cover the whole wavelength range between $\sim$\,2000 and $\sim$\,11\,000\,\AA\ for the first time since day 8329 after the explosion. We studied both the broad- and narrow-band photometry and morphology of the ER, the ejecta, and the central region at this new epoch and provided comparisons with the previous one. We also analyzed the light curves of the supernova in the F438W and F625W filters and added four more epochs to the latest light curves provided in the literature. 

The ER has followed its slow fading route over the whole wavelength range covered by our observations (see Fig.\,\ref{fig:ratio_SEDs_2009_and_2022}) and shows a flux at day 12\,980 about half of the flux at day 8329. The eastern part has faded more than the western part by about $10-20$\%. We also find a reversal in the time evolution of the ratio between western and eastern parts fluxes in the ER: While this ratio increased until $\sim 11\,000$\,days after the explosion, it started to decrease afterwards, though the western part is still a factor of two brighter than its eastern counterpart. 
 
The ejecta showed a light curve that started to flatten around day 11\,000 after the explosion. Overall, the ejecta have proportionally brightened less between epochs 8329 and 12\,980 (by a factor $1.1-1.3$ in $\sim4650$ days) than between epochs 5400 and 7200 (where it brightened by a factor 2.6 in $\sim1800$ days). This can be attributed to less energy input from X-rays from the ER at late times; the soft X-ray light curve declines after day $\sim10\,000$ and the hard X-ray light curve flattens since day $\sim12\,000$ after the explosion. The western part of the ejecta is brighter than the eastern part, consistent with the western ER being brighter in the X-rays than the eastern part. The ejecta have brightened less in the F814W filter than in the other broad-band filters because the emission in this filter comes principally from [Ca\,\textsc{ii}] which is likely partly excited by the fading optical emission from the ER.

The emission in the ejecta follows homologous expansion. The morphology is clearly asymmetric and the ejecta are well-mixed on large scales, as evident from the very similar morphology observed in all filters, which probe emission lines from different elements, including H, Fe, and Ca. Only one broad-band filter stands out from this picture: F275W. The difference in morphology observed in the F275W filter is attributed to the different mechanism producing the Mg\,\textsc{ii} $\lambda\lambda$\,2795, 2802 lines, as well as the fact that these lines are optically thick. These lines are partly formed through pumping by the Ly\,$\alpha$ and, to a smaller extent, the N\,\textsc{v} $\lambda\lambda$\,1239, 1243 photons from the reverse shock, as well as by thermalization of X-rays coming from the ER. 

Finally, we searched for emission associated with the compact object in view of recent JWST observations of SN 1987A in the NIR and mid-IR that show a central source. However, the HST observations do not show any sign of a compact object, neither in the morphology, nor in the photometry, nor in the light curves. The models discussed in the context of the JWST observations predict a relatively strong [O\,\textsc{iii}] $\lambda$\,5008 line. We used the F502N observation containing this line to put a $3\sigma$ upper limit on a point source of  $6.1\times10^{-18}$\,erg\,s$^{-1}$\,cm$^{-2}$. This value being lower than both CNS and PWN model predictions, it does not favor the CNS nor the PWN as the ionizing source. However, there are several caveats in the model predictions and the current observations are fully consistent with the CNS model in the scenario that dust scattering spreads the emission over a larger region. The non-detection of the [O\,\textsc{iii}] $\lambda$\,5008 line can also be an indication that the emission seen with JWST originates from the inner part of the S and Ar rich zones. 

In the future, we hope to continue the monitoring in F502N and other filters to try to detect the presence of the compact object. The probability of detecting optical emission associated with the compact object increases with time as the optical depth of the dust decreases. 

The reverse shocks will be addressed in a forthcoming paper: Their properties will be determined through the analysis of the time evolution and modelling of the UV spectrum. This will allow us to gain information about the shock physics and the properties of the circumstellar medium, the latter probing the mass-loss history of the progenitor including the formation of the rings.

\begin{acknowledgments}
\textit{Acknowledgments.} S. Rosu warmly thank Prof. G. Rauw and Dr. D. Alp for constructive discussions. Support for HST GO program numbers 11653, 12241, 13181, 13405, 13810, 14333, 14753, 15256, 15503, 15928, 16265, and 16789 was provided by NASA through grants from the Space Telescope Science Institute, which is operated by the Association of Universities for Research in Astronomy, Inc., under NASA contract NAS5-26555. M. Matsuura acknowledges support from  STFC Consolidated grant (ST/W000830/1).
\end{acknowledgments}

Data presented in this article were obtained from the Mikulski Archive for Space Telescopes (MAST) at the Space Telescope Science Institute. The specific observations analyzed can be accessed via\dataset[DOI: 10.17909/vv4m-mm08]{https://doi.org/10.17909/vv4m-mm08}

\vspace{5mm}
\facilities{HST (WFC3)}
\software{astropy \citep{astropy13, astropy18, astropy22}, DrizzlePac \citep{ferland13}, matplotlib \citep{bertin96}, Photutils \citep{bradley23}, scipy \citep{virtanen20}}

\appendix
\section{Additional figures\label{appendix:additionalfigures}}
\begin{figure*}
\centering
\includegraphics[clip=true,trim=0 60 0 0,width=\linewidth]{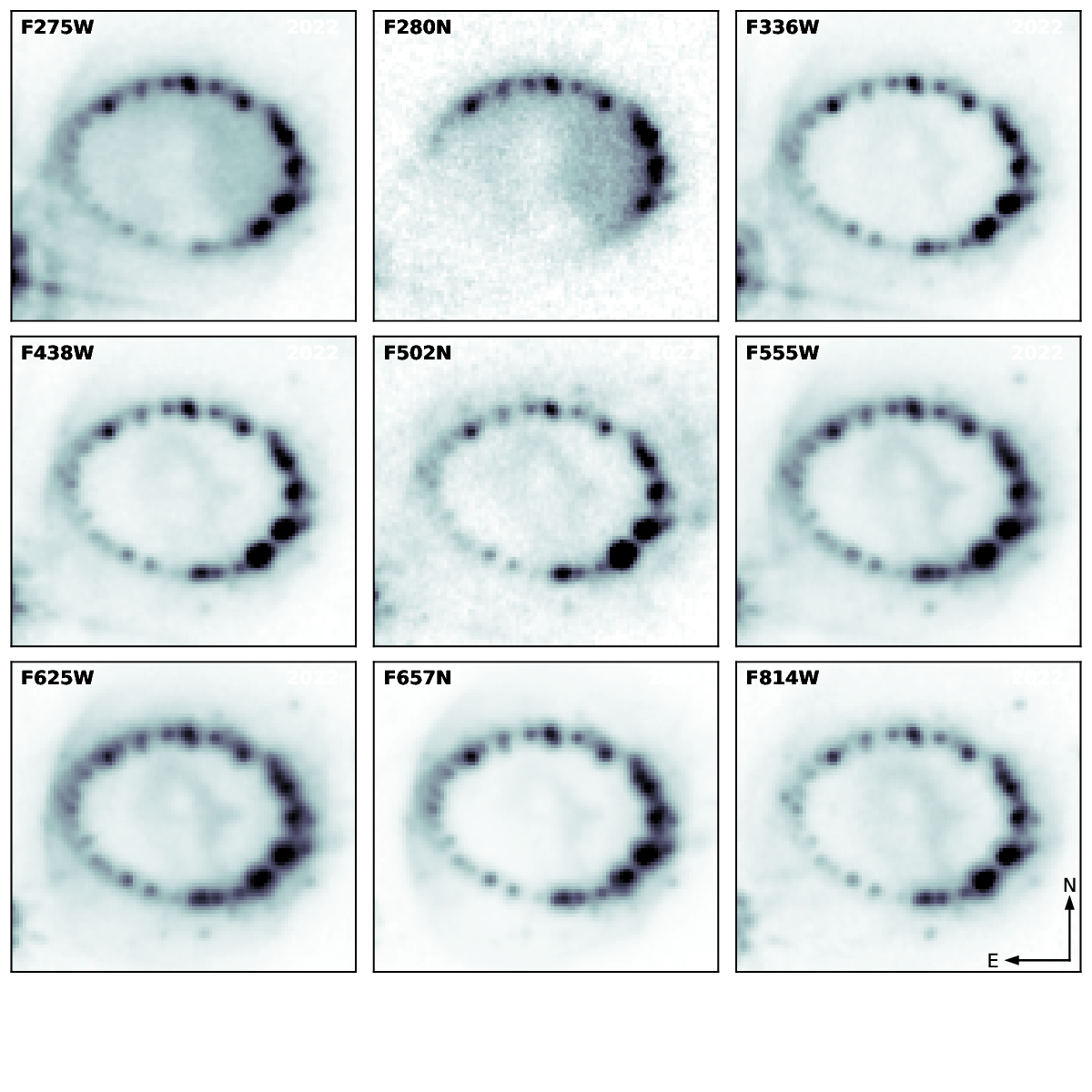}
\caption{HST/WFC3 images of SN\,1987A taken 12\,980 days after the explosion in nine different filters. The images were scaled by an $\asinh$ function and the color scales were chosen differently for each filter to highlight the hotspots in the ER. The field of view for each  image is $2\fds50 \times 2\fds25$.\label{fig:ER_all_2022}}
\end{figure*}

\begin{figure*}
\centering
\includegraphics[clip=true,trim=0 60 0 0,width=\linewidth]{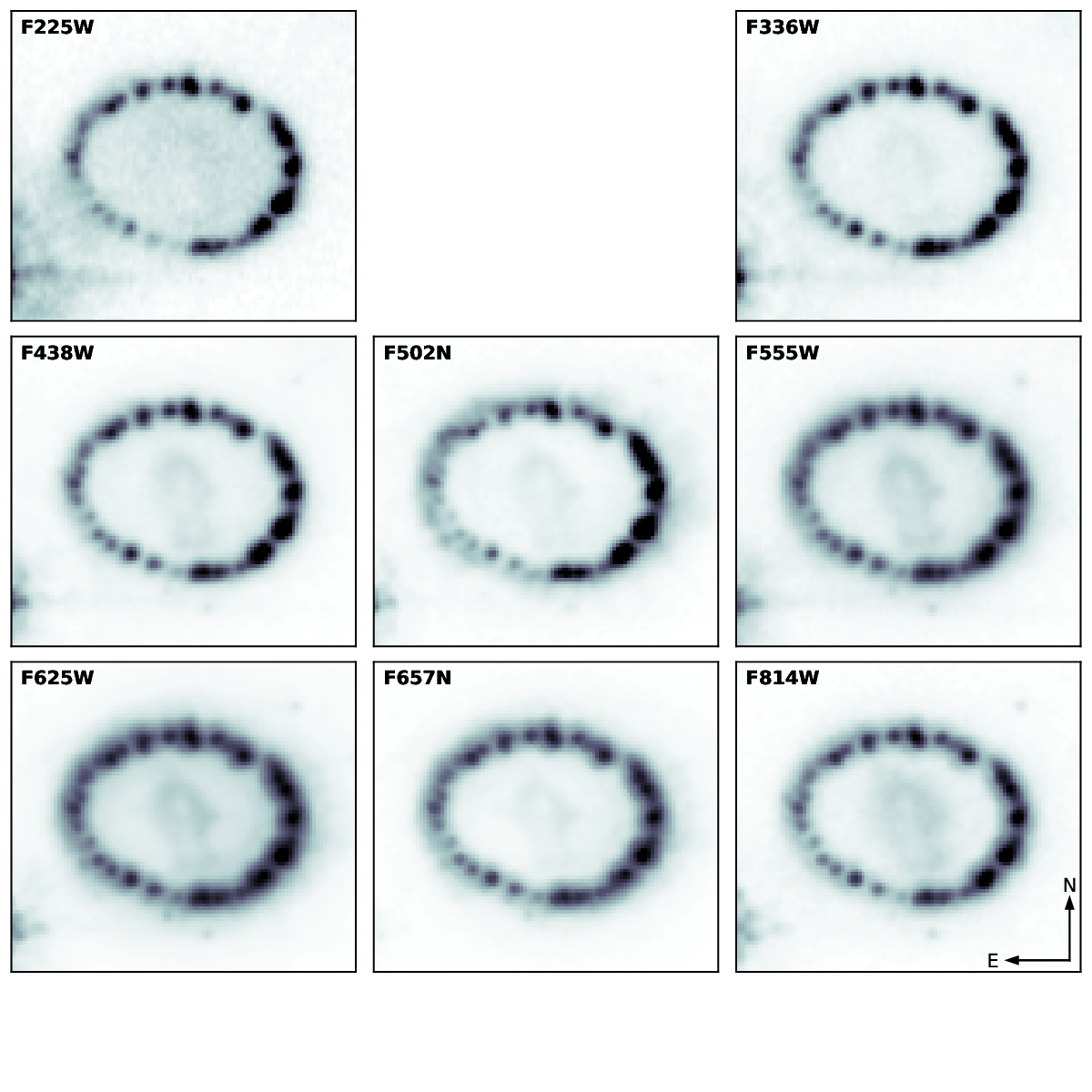}
\caption{HST/WFC3 images of SN\,1987A taken 8329 days after the explosion in eight different filters. The images were scaled by an $\asinh$ function and the color scales were chosen differently for each filter to highlight the hotspots in the ER. The field of view for each  image is $2\fds50 \times 2\fds25$.\label{fig:ER_all_2009}}
\end{figure*}

\begin{figure*}
\centering
\includegraphics[clip=true,trim=0 0 0 190,width=1\linewidth]{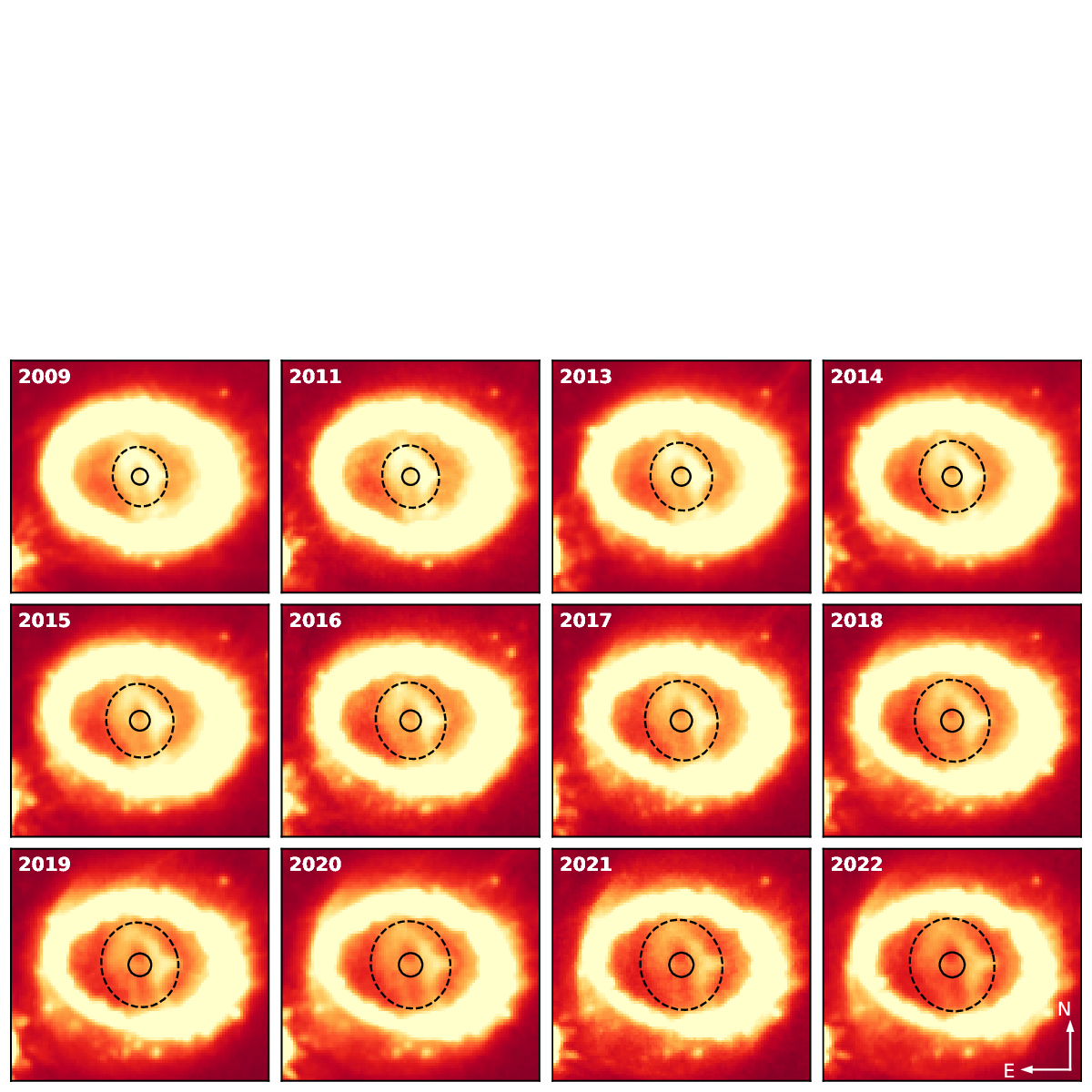}
\caption{HST/WFC3 images showing the evolution of SN\,1987A in the F625W filter between epochs 8329 and 12\,980 (labeled by the year of observation). The emission in the lower left corner is due to Star\,3 (see Fig.\,\ref{fig:general}). The images were scaled by an $\asinh$ function to highlight the weak emission in the ejecta. The field of view for each image is $2\fds50 \times 2\fds25$. The black circle and dashed black ellipse overplotted on each image represent the center and ejecta regions, respectively,  adopted to compute the fluxes. These regions grow in size from 2009 to 2022, using Eq.\,\eqref{eq:ejectavel} to define a co-moving volume that tracks the same expanding ejecta with time.} \label{fig:Ejecta_f625w_2009_to_2022}
\end{figure*}

\begin{figure*}
\centering
\includegraphics[clip=true,trim=0 0 0 190,width=0.9\linewidth]{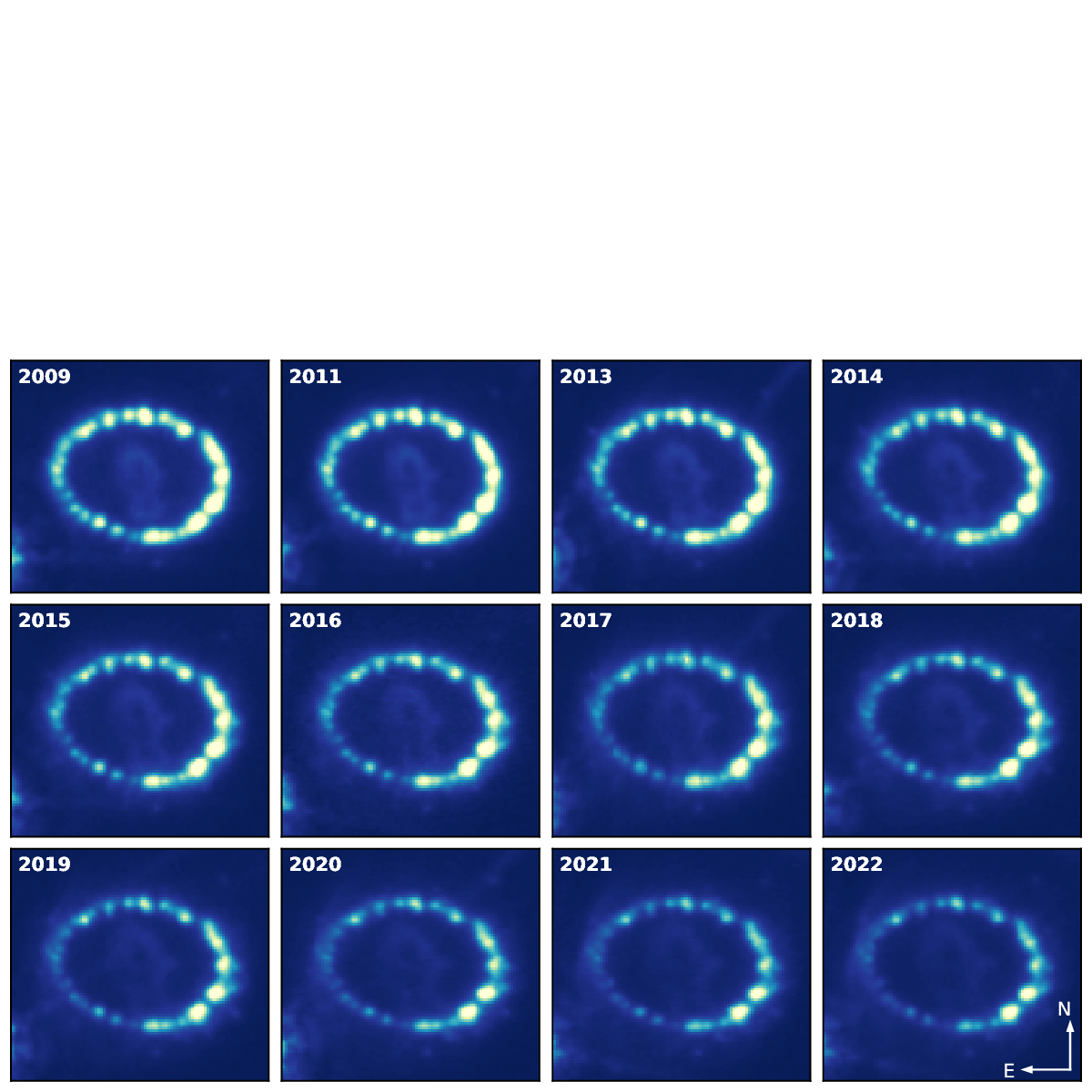}
\includegraphics[clip=true,trim=0 0 0 190,width=0.9\linewidth]{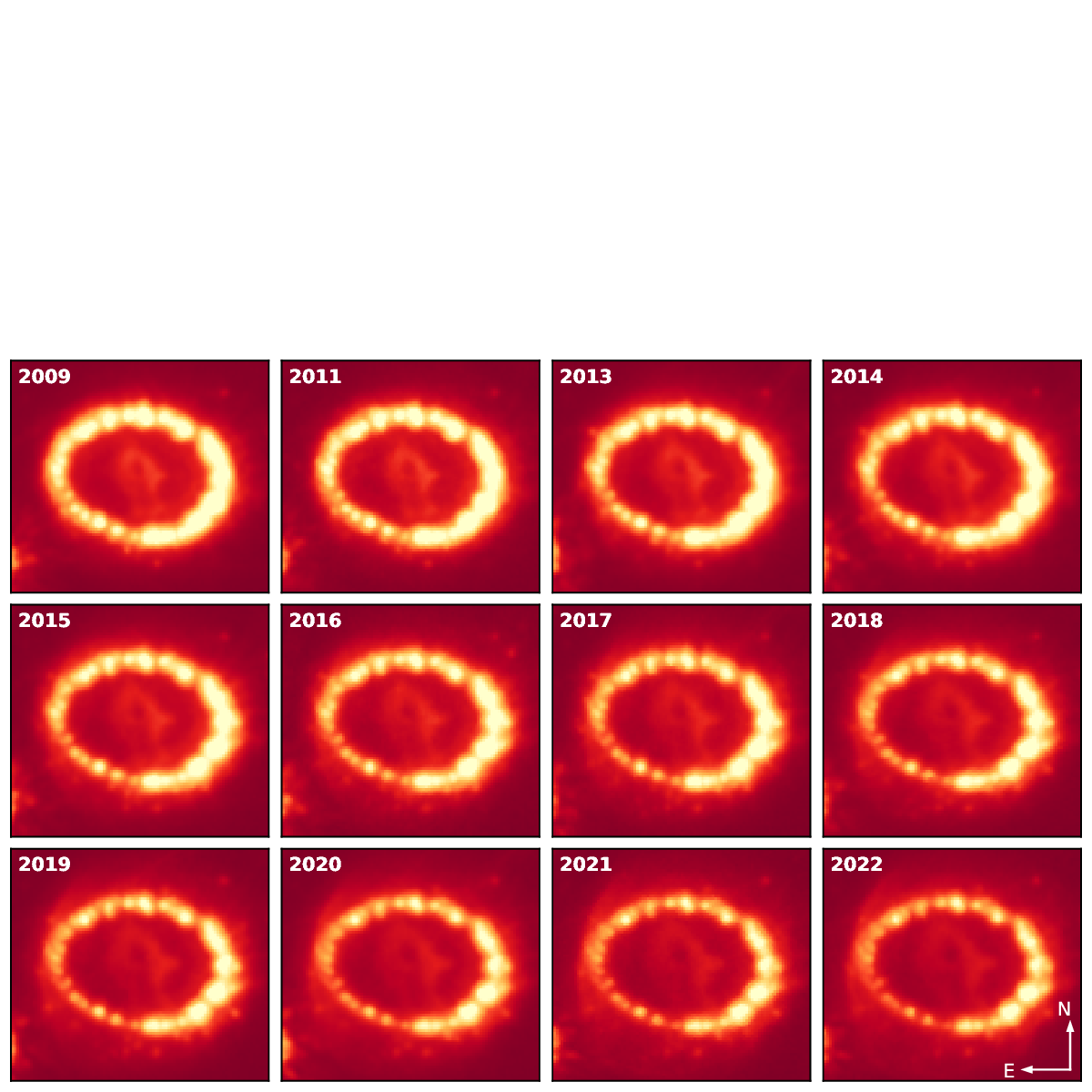}
\caption{HST/WFC3 images showing the evolution of SN\,1987A in the F438W (top 3 rows) and F625W (bottom 3 rows) filters between epochs 8329 and 12\,980 (labeled by the year of observation). The emission in the lower left corner is due to Star\,3 (see Fig.\,\ref{fig:general}). The images were scaled by an $\asinh$ function. The field of view for each image is $2\fds50 \times 2\fds25$. \label{fig:ER_all_2009_to_2022}}
\end{figure*}

This section presents the HST observations at days 12\,980 (Fig.\,\ref{fig:ER_all_2022}) and 8329 (Fig.\,\ref{fig:ER_all_2009}) in all filters with a color scale optimized to highlight the hotspots in the ER and between days 8329 and 12\,980 in the F625W filter with a color scale optimized to highlight the emission in the ejecta (Fig.\,\ref{fig:Ejecta_f625w_2009_to_2022}) and in the F438W and F625W filters (Fig.\,\ref{fig:ER_all_2009_to_2022}) with a color scale optimized to highlight the hotspots in the ER. 

\section{ER contribution to the ejecta and center\label{appendix:ER_synthetic}}
In this appendix, we describe how the contribution of the ER to the ejecta and center was removed for all observations. The ER is indeed more than an order of magnitude brighter than the ejecta, which means that the scattered light from the ER -- the extended wings of the PSFs of the hotspots -- significantly contributes to the background in the ejecta.

We constructed a synthetic ER aiming at reproducing the ER for each epoch based on the locations of the 28 bright spots in the ER\footnote{We note that we only consider the hotspots in the main ring, not new spots appearing further out at later times.} \citep{tegkelidis23}. We used the \texttt{Tiny Tim} HST PSF modeling tool to generate PSFs at the locations of the spots which amplitudes are given by the relative brightnesses of the spots \citep{tegkelidis23}. We adopted the corresponding filters and the 2017 spectra of the ER \citep{kangas22} to generate the PSFs. The total flux of the synthetic ER was then scaled to match the observations. We here stress that although the synthetic ER does not perfectly reproduce the observed ER, it does not significantly affect the correction for ER light contributing to the ejecta apertures as long as point sources with the correct total flux are placed in an ellipse, as discussed in \citet{larsson19a}.  

Based on these synthetic ERs, we computed the total contribution of the ER to the corresponding ejecta and center regions. We removed these contributions from the total ejecta and center count rates. We found that the fluxes were reduced by a non-negligible amount, as summarized in Table\,\ref{table:flux_reduction_2022_2009} for the epochs 12\,980 and 8329 observations and in Table\,\ref{table:flux_reduction_2011_2021} for the observations from epoch 8717 to 12\,598 included.

\begin{table}[h]
\centering
\caption{Reduction of the flux (in \%) in the ejecta and center regions after removal of the ER contribution in the images at epochs 12\,980 and 8329 in the different filters.}
\begin{tabular}{lllll}
\hline\hline
Filter & \multicolumn{2}{c}{Epoch 12\,980 days} & \multicolumn{2}{c}{Epoch 8329 days} \\
& Ejecta & Center & Ejecta & Center \\
\hline
F225W & ... & ... & 10.0& 9.1 \\
F275W & 9.7 & 8.8 & ... & ... \\
F280N & 7.4 & 7.8 & ... & ... \\
F336W & 16.5 & 14.8 & 16.7 & 15.7 \\
F438W & 12.6 & 11.8 & 10.6 & 11.0 \\
F502N & 10.7 & 10.5 & 12.3 & 13.9 \\
F555W & 12.5 & 9.4 & 10.6 & 9.3 \\
F625W & 14.1 & 9.6 & 12.5 & 9.6 \\
F657N & 20.8 & 14.0 & 19.1 & 14.3 \\
F814W & 9.1 & 6.7 & 7.0 & 6.3 \\
\hline
\end{tabular}
\label{table:flux_reduction_2022_2009}
\end{table} 

\begin{table}
\centering
\caption{Reduction of the flux (in \%) in the ejecta and center regions after removal of the ER contribution in the F438W and F625W filters in the images between epochs 8717 and 12\,598.}
\begin{tabular}{lllll}
\hline\hline
Epoch (days) & \multicolumn{2}{l}{Filter F438W} & \multicolumn{2}{l}{Filter F625W} \\
& Ejecta & Center & Ejecta & Center \\
\hline
8717 & 11.2 & 11.9 & 12.6 & 9.8 \\
9480 & 11.7 & 11.8 & 12.9 & 9.6 \\
9974 & 12.0 & 12.1 & 13.2 &9.7 \\
10\,317 & 12.1 & 12.6 & 13.3 & 9.9 \\
10\,698 & 12.0 & 12.0 & 13.4 & 9.6 \\
11\,119 & 11.9 & 11.7 & 13.2 & 9.4 \\
11\,458 & 12.4 & 12.0 & 13.8 & 9.7 \\
11\,837 & 12.5 & 11.6 & 14.0 & 9.7\\
12\,218 & 12.3 & 11.3 & 13.9 & 9.2\\
12\,598 & 12.3 & 11.6 & 14.2 & 9.7\\
\hline
\end{tabular}
\label{table:flux_reduction_2011_2021}
\end{table} 

\section{Outer ring contribution to the ejecta and center}
\label{appendix:outerrings}
In this appendix, we provide the estimated contribution of the northern outer ring to the ejecta and center regions at epochs 12\,980 and 8329 in the different filters (see Table\,\ref{table:outerring}). These contribution were estimated using the brightest and faintest parts of the northern outer ring (northeast and southwest) as minimum and maximum fluxes, adopting the same number of pixels as in the ejecta and center regions.
It is expected that the contribution from the northern outer ring lies in the ranges provided in Table\,\ref{table:outerring}, but we note that we did not correct the computed fluxes for it since we do not know the exact values of the contributions. Given that the dust in the central ejecta may scatter and absorb light from the outer ring, the actual outer ring contribution may be even lower than these limits.

\begin{table}
\centering
\caption{Estimated contribution (in \%) of the northern outer ring to the ejecta and center regions at epochs 12\,980 and 8329 in the different filters.}
\begin{tabular}{lllll}
\hline\hline
Filter & \multicolumn{2}{c}{Epoch 12\,980 days} & \multicolumn{2}{c}{Epoch 8329 days} \\
& Ejecta & Center & Ejecta & Center \\
\hline
F225W & ... & ... & 0--5 & 1  \\
F275W & 1--2 & 5--8 & ... & ... \\
F280N & 0--1 & 0--2 & ... & ... \\
F336W & 2--6 & 8--19 & 1--9 & 3--18 \\
F438W & 1--5 & 5--17 & 0--7 & 1-15 \\
F502N & 4--13 & 15--46 & 3--11 & 8--25 \\
F555W & 2--4 & 6--13 & 1--6 & 2--12 \\
F625W & 2--3 & 5--10 & 1--4 & 3--8 \\
F657N & 2--5 & 7--15 & 2--5 & 4--10 \\
F657N & 2--5 & 7--15 & 2--5 & 4--10 \\
F814W & 1 & 3--4 & 1 & 1--3 \\
\hline
\end{tabular}
\label{table:outerring}
\end{table} 

\section{Diffraction spike contributions\label{appendix:diffraction_spikes}}
In this appendix, we describe how the diffraction spike contributions to the ejecta, center, and ER regions were removed. The center region is affected by a diffraction spike from Star 2 at epochs 11\,119 and 12\,218 (see Figs\,\ref{fig:center_f438w_2009_to_2022} and \ref{fig:Ejecta_f625w_2009_to_2022}). The ejecta region is affected by a diffraction spike from Star 2 at epochs 9480, 11\,119, and 12\,218, and from Star 3 at epochs 8717, 11\,458, and 11\,837 (see Figs\,\ref{fig:center_f438w_2009_to_2022} and \ref{fig:Ejecta_f625w_2009_to_2022}). The ER region is affected by a diffraction spike from Star 2 at epochs 9480, 11\,119, 11\,837, and 12\,218, and from Star 3 at epochs 8717, 9480, 9974, 10\,317, 11\,119, 11\,458, 11\,837, and 12\,218 (see Figs\,\ref{fig:center_f438w_2009_to_2022} and \ref{fig:Ejecta_f625w_2009_to_2022}). 

\begin{figure}
\centering
\includegraphics[clip=true, trim=100 5 100 185,width=0.5\linewidth]{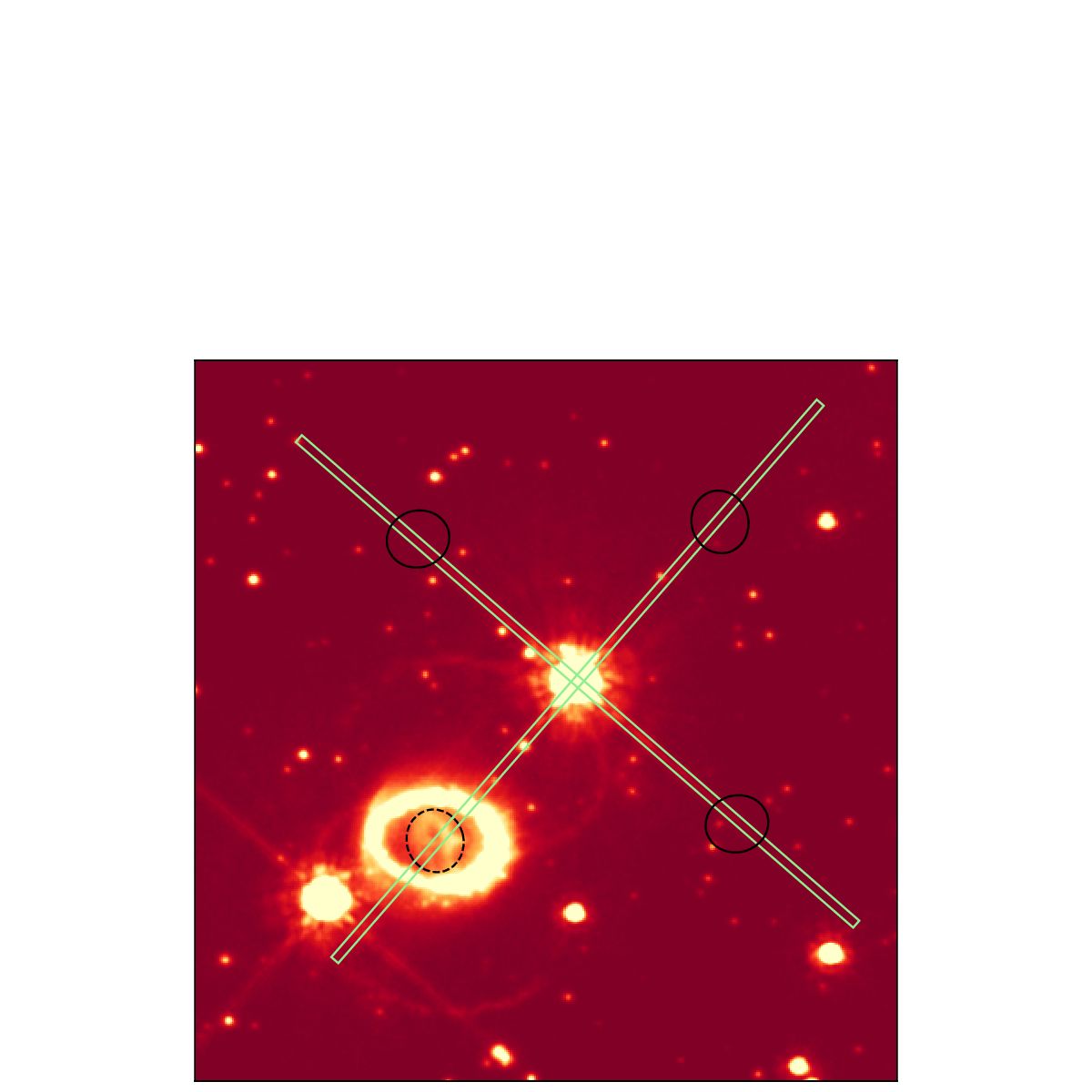}
\caption{HST/WFC3 image of SN\,1987A at day 12\,218 in the F625W filter illustrating how the diffraction spike contribution (from Star 2 in this case) was estimated. The green rectangles represent the diffraction spikes regions, the dashed black ellipse represents the ejecta region, and the plain black ellipses represent the ejecta region rotated by $90^\circ$, $180^\circ$, and $270^\circ$. The image was scaled by an $\asinh$ function. The field of view is $9\fds50 \times 9\fds75$.}
\label{fig:diffraction_spikes}
\end{figure}

To account for these artefacts, we computed the fluxes in the three intersection regions between the three diffraction spikes that do not cross the center, ejecta, or ER region, and the corresponding center, ejecta, or ER region rotated by $90^\circ$, $180^\circ$, and $270^\circ$ with respect to the center of the considered Star 2 or 3 (see an illustration for the observation at day 12\,218 in the F625W filter in Fig.\,\ref{fig:diffraction_spikes}). We adopted a five-pixels width for the diffraction spikes. For each observation and each considered region, we then averaged the computed fluxes to get an estimate of the diffraction spike contribution to the region. We discarded the diffraction spikes that were contaminated by either a star in the field or the outer ring of SN\,1987A. The reduction of the fluxes after subtraction of the diffraction spike contributions to the ejecta, center, and ER regions are given in Table\,\ref{table:diffraction_spikes}.

\begin{table*}
\caption{Reduction of the flux (in \%) in the ejecta, center, and ER regions after removal of the diffraction spike contributions from Stars 2 and 3 in the F438W and F625W filters in the images between epochs 8717 and 12\,218.}
\begin{tabular}{llllllllll}
\hline\hline
Epoch (days) & \multicolumn{3}{c}{Ejecta} & \multicolumn{3}{c}{Center} & \multicolumn{3}{c}{ER} \\
& F438W & F625W & Star & F438W & F625W & Star & F438W & F625W & Star \\
\hline
8717 & 2.25 & 0.83& 3 & ... & ... & ... & 0.29& 0.09& 3 \\
9480 &  0.59 & 0.56& 2 & ... & ... & ... & 0.80& 0.24& 2 \& 3 \\
9974 & ... & ... & ... & ... & ... & ... & 0.17& 0.05& 3  \\
10\,317 & ... & ... & ... & ... & ... & ... & 0.16& 0.05& 3\\
11\,119 & 1.18 & 1.06& 2 & 2.72 & 2.24 & 2 & 0.67& 0.23& 2 \& 3\\
11\,458 & 3.13 & 1.10& 3 & ... & ... & ... & 0.36& 0.11& 3 \\
11\,837 & 3.91 & 1.25& 3 & ... & ... & ... & 0.54& 0.23& 2 \& 3  \\
12\,218 & 1.34 & 1.22& 2 & 4.58& 3.92& 2 & 1.42& 0.46& 2 \& 3 \\
\hline
\end{tabular}
\label{table:diffraction_spikes}
\end{table*} 

\end{document}